\documentclass[twocolumn]{aastex63}

\usepackage{cancel}

\usepackage{amsmath}
\usepackage{graphicx}
\usepackage{float}
\usepackage{listings}
\usepackage{cancel}
\usepackage{url}
\usepackage{natbib}
\usepackage{wrapfig}
\usepackage{nicefrac}
\usepackage{placeins}

\newcommand{\pderiv}[2]{\frac{\partial #1}{\partial #2}}

\newcommand{\deriv}[2]{\frac{d #1}{d #2}}

\accepted{January 23, 2021}

\shorttitle{Line Driven Winds}
\begin{document}

\title{An Updated Formalism For Line-Driven Radiative Acceleration and
Implications for Stellar Mass Loss
}

\correspondingauthor{Aylecia S. Lattimer}

\author[0000-0002-2004-5084]{Aylecia S. Lattimer}
\affiliation{Department of Astrophysical and Planetary Sciences,
Laboratory for Atmospheric and Space Physics,
University of Colorado, Boulder, CO 80309, USA}

\author[0000-0002-3699-3134]{Steven R. Cranmer}
\affiliation{Department of Astrophysical and Planetary Sciences,
Laboratory for Atmospheric and Space Physics,
University of Colorado, Boulder, CO 80309, USA}

\begin{abstract}
Radiation contributes to the acceleration of large-scale flows in various astrophysical environments because of the strong opacity in spectral lines. Quantification of the associated force is crucial to understanding these line-driven flows, and a large number of lines (due to the full set of elements and ionization stages) must be taken into account. Here we provide new calculations of the dimensionless line strengths and associated opacity-dependent force multipliers for an updated list of approximately 4.5 million spectral lines compiled from the NIST, CHIANTI, CMFGEN, and TOPbase databases. To maintain generality of application to different  environments, we assume local thermodynamic equilibrium, illumination by a Planck function, and the Sobolev approximation. We compute the line forces in a two-dimensional grid of temperatures (i.e., values between 5,200 and 70,000 K) and densities (varying over 11 orders of magnitude). Historically, the force multiplier function has been described by a power-law function of optical depth. We revisit this assumption by fitting alternative functions that include a saturation to a constant value (Gayley's $\bar{Q}$ parameter) in the optically-thin limit. This alternate form is a better fit than the power-law form, and we use it to calculate example mass-loss rates for massive main-sequence stars. Because the power-law force multiplier does not continue to arbitrarily small optical depths, we find a sharp decrease, or quenching, of line-driven winds for stars with effective temperatures less than about 15,000 K.

\end{abstract}

\keywords{Stellar winds (1636), Stellar mass loss (1613), Atomic Physics (2063), Radiative Processes (2055)}

\section{Introduction} \label{sec:intro}
The study of flows driven by photons has a long history. For example, \cite{Johnson1925} hypothesized that that helium atoms could be ejected from a star via radiation pressure, suggesting that photon-driven flows can be responsible for stellar mass loss. We now know that photon-driven outflows exist in various astrophysical environments, including accreting objects such as protostars, cataclysmic variables, and active galactic nuclei (AGN) \citep{Owocki2004,Puls2008,Higginbottom2014}. The driving mechanism of these flows by photons is often referred to as ``radiation pressure,'' where the force of radiation from the spectral lines acts on the material of the outflow. The very large number of spectral lines in any given ion has a dominant effect on this pressure on the flow material \citep{Castor1974,Castor1975}. The absorption and re-emission of photons in a spectral line of frequency $\nu_0$ results in a radial transfer of momentum, driving a wind outward from the star.

Quantifying and understanding mass and energy flows is critical in the understanding of how affected objects evolve and interact with their surroundings. Massive star winds can be directly observed in the stellar spectral energy distributions once the star is above a certain threshold luminosity \citep{Kudritzki&Puls2000}. Ultraviolet (UV) observations showing P Cygni profiles and high flow velocities initially suggested that an extension of solar wind theory was insufficient to explain the winds of O and B stars \citep{Cassinelli1979}. \cite{Lucy&Solomon1970} demonstrated that the absorption and re-emission of photons by 12 UV resonance lines can drive a wind in O type stars, consequently driving mass loss in the form of the stellar wind. 
\cite{Castor1975} (hereafter CAK) found that the large number of lines in any given ion of an atom contribute to the radiation force, proportional to the continuum flux at their specific frequency. Using 900 multiplets of C III, CAK found a mass loss rate estimate $\sim$100 times greater than those predicted by \cite{Lucy&Solomon1970}, leading to the conclusion that the force from the lines should exceed gravity by approximately two orders of magnitude for small optical depths in O type stars. Therefore, O stars cannot have static atmospheres, as there is no mechanism that can prevent the ejection of the surface layers of the star. This was a major advancement in the understanding of stellar winds. Models based on the theory of line-driven winds have yielded results that agree well with observations of mass loss and terminal flow rates \citep{Owocki1988,Puls2008,Sundqvist2019}.  

Line-driven outflows are also encountered in various other environments. For example, broad absorption lines are present in the UV spectra of quasars, as well as in other wavelengths. In these cases, the blueshift of the lines suggests the presence of winds from the AGN, sometimes with velocities of up to $0.2c$, making a line-driven disk wind a promising hydrodynamical scenario for AGN outflows \citep{Proga2007, Risaliti&Elvis2010}. Additionally, \cite{Kee2016} suggested that the strong line-driven winds from OB type stars with circumstellar disks could drive ablation of the disk's surface layers on short timescales that could be a contributing factor to the relative rarity of O type stars in the galaxy (see also \citealp{disk_ablation_2_2018}, \citealp{disk_ablation_3_2018}, \citealp{disk_ablation_4_2019}). By themselves, winds from massive stars can drive galactic evolution by injecting momentum, energy, and stellar material into the interstellar medium \citep{Kudritzki&Puls2000}.

Previous authors have computed lists of spectral lines and modeled the distributions of the line strength parameters (see for example \citealp{Abbott1982, Shimada1994, Gormaz-Matamala2019}). However, these line lists have been comprised of significantly fewer lines than those currently available. \citet{Lucy&Solomon1970} originally used 12 lines, CAK updated this to include 900 transitions, and \citet{Pauldrach1987} used a list of 10,000 transitions. More recently, \citet{Gormaz-Matamala2019} used a list of $\sim$900,000 lines, and \citet{Bjorklund2020} has used a database of approximately one million lines (see also \citealt{Sundqvist2019}). In this work we construct a newly updated line list, comprised of 4,514,900 spectral lines. To this end, atomic data is assembled from four separate databases. We also reexamine the validity of the historical assumption of a power law to describe the line strength distribution.  

This work primarily aims to provide new insight into the form of the line-force multiplier, beyond that of the CAK power-law form. This new form will also provide an alternative to computationally expensive simulations that use the full line list, such as those described in \citet{Gormaz-Matamala2019}. To do this, we make some basic assumptions, such as those of local thermodynamic equilibrium (LTE), and an initial Planck function central source. These assumptions and the calculation process of dimensionless line strengths and weighting functions are described in further detail in Section \ref{sec:Radiative Acceleration}. Section \ref{sec:Atomic Data} describes the collection and compilation of the spectral lines. Section \ref{sec:Force Multiplier} describes the calculation of the updated force multiplier $M(t)$, as well as the fitting of a CAK-form power-law (Section \ref{subsec: CAK fits}) and an alternate function (Section \ref{subsec: alt fit function}) to the calculated values. Section \ref{sec: Mass-Loss Rates} describes the calculation of mass-loss rates from both the CAK (Section \ref{subsec: CAK mass losses}) and alternate (Section \ref{subsec: alt mass losses}) multiplier forms, and a comparison of the two (Section \ref{subsec: mass loss comparison}). We end with a discussion of our conclusions in Section \ref{sec:Conclusions}.

\section{Radiative Acceleration of Line-Driven Flows} \label{sec:Radiative Acceleration}
A general way of expressing the radiative acceleration ${\bf g}_{\rm rad}$ on a parcel of gas is to take the first moment of the equation of radiative transfer.  Following \citet{HubenyMihalas2015},
\begin{equation}
  {\bf g}_{\rm rad} ({\bf r}) \, = \, \frac{1}{c}
  \int d\nu \int d\Omega \hat{\bf n} \,
  ( \kappa_{\nu} I_{\nu} - j_{\nu} )
\end{equation}
where $\hat{\bf n}$ is the unit vector specifying an arbitrary ray-path, $\kappa_{\nu}$ and $j_{\nu}$ are the absorption coefficient (cm$^2$ g$^{-1}$) and emissivity, and $I_{\nu}$ is the specific intensity. In the comoving frame of a flow, it is often assumed that $j_{\nu}$ is an even function of $\hat{\bf n}$ (so it cancels out of the above moment integral) and that angle anisotropies of $\kappa_{\nu}$ are sufficiently weak to allow it to be taken out of the solid-angle integral. Thus, the expression we use is
\begin{equation} \label{eq: rad acceleration}
    \textbf{g}_{\rm rad}(\textbf{r}) = \frac{1}{c}\int \kappa_\nu \textbf{F}_\nu(\textbf{r}) d\nu
\end{equation}
where ${\bf F}_{\nu}$ is the radiative flux (photon energy flux). Below, we also write the opacity $\chi_\nu$ in units of cm$^{-1}$ as 
\begin{equation} \label{eq: opacity defs}
    \chi_\nu = \kappa_\nu\rho = \sigma_\nu n
\end{equation}
where it is sometimes useful to use the absorption coefficient $\kappa_{\nu}$
or the cross section $\sigma_{\nu}$, and we also define the mass density $\rho$ and number density $n$ in units of cm$^{-3}$.

From Equation (\ref{eq: rad acceleration}), we see that opacity is necessary for any acceleration due to the radiation field to occur; with zero opacity, the radiation cannot interact with the material of the flow. Spectral lines have a dramatic effect on the wind driving due to resonance that occurs when encountering continuum photons, leading to a large cross section. This effect can be strong enough that it is still magnified after being averaged over the entire continuum. Therefore, to find the total radiative force on a parcel of gas of given temperature and density, all the spectral lines that are encountered by the radiation field must be accounted for. The inclusion of as many lines as possible is imperative in developing a more complete understanding of this phenomenon.

\subsection{Dimensionless Line Strengths} \label{subsec:q_i}
We follow \citet{Gayley1995} in characterizing the distribution of spectral line strengths as a set of dimensionless ratios $q_i$ that describe the atomic physics, and dimensionless weighting factors $\widetilde{W}_i$ that describe the illumination of the atom from
a given spectral energy distribution. The product $q_i \widetilde{W}_i$ represents the full ratio of radiative acceleration due to a specific line $i$ to the acceleration on free electrons. In such a ratio of accelerations (see Equation (\ref{eq: rad acceleration})), the pre-factors of $1/c$ cancel out, and we choose to multiply both the numerator and denominator by the mass density $\rho$. We can write the bound line opacity as 
\begin{equation}
    \chi_\nu = \chi_{\rm{L}} \phi(\nu)
\end{equation}
where $\phi(\nu)$ is the line profile function.
Thus, the ratio of accelerations can be written as
\begin{equation}
    \frac{g_{\rm bound}}{g_{\rm free}} = \frac{\int \chi_\nu F_\nu d\nu}{\int \chi_e  F_\nu d\nu} = \frac{\chi_L \int \phi(\nu)F_\nu d\nu}{\chi_e \int F_\nu d\nu}.
\end{equation}
For this work, we assume that the environments in question are in LTE. This assumption (also used by CAK) is often set aside when modeling line-driven winds from, e.g., massive stars, but here we retain it to maintain a level of generality concerning the environments in question. Of course, this assumption will need to be reevaluated in future applications to specific systems \cite[see, e.g,][]{Pauldrach1987}.
Under the assumption of LTE, the quantity $\chi_L$ is then given by
\begin{equation}\label{eq:chi_L}
    \chi_L = \frac{\pi e^2}{m_e c}f_{ij}n_i\left(1-e^{-h\nu_0/kT}\right),
\end{equation}
where $f_{ij}$ is the semiclassical oscillator strength, where $i$ and $j$ are used to refer to the lower and upper atomic levels, respectively.
The Thomson scattering opacity $\chi_e$ is given by 
\begin{equation}\label{eq:Thomson opacity}
    \chi_e = \sigma_T n_e = \left(\frac{8\pi r_e^2}{3}\right)n_e
\end{equation}
where $\sigma_T$ is the Thomson scattering cross section and the classical electron radius is given by 
\begin{equation}
    r_e = \frac{e^2}{m_e c^2}.
\end{equation}

Each line profile function $\phi\nu)$ is very narrow when integrated over the energy
distribution, so we can approximate it to a Dirac delta function when evaluated at frequency $\nu_0$
\begin{equation}
    \int \phi(\nu)F_\nu d\nu \approx F_{\nu_0}.
\end{equation}
We can additionally define the frequency integrated flux $F$ as 
\begin{equation}
    \int F_\nu d\nu \equiv F.
\end{equation}
We can then define a dimensionless weighting ratio
\begin{equation}\label{eq:W_i}
    \widetilde{W}_i = \frac{\nu_0 F_{\nu_0}}{F},
\end{equation}
which accounts for the flux integrals in Equation (3).

In this paper, we maintain a generality of application by assuming a Planck function for the illuminating flux. We use a temperature $T$, presumed equal to the local temperature of the gas ($T=T_{\rm rad}=T_{\rm eff}$), to characterize this Planck function \citep{Noebauer+Sim2015,Gormaz-Matamala2019}. Although some authors have suggested the electron temperature is a fraction (usually 0.8$T_{\rm eff}$) of the effective temperature, the radiative temperature is often taken to be equal to that of $T_{\rm eff}$ in the Planck case \citep[see, for example,][]{Puls2000}.   
This assumption is commonly used for the wind-driving circumstellar regions near massive-star photospheres (e.g., \citealt{Drew1989}, \citealt{Kee2016}), in which radiative cooling in an isothermal gas efficiently maintains a nearly constant temperature $T$, roughly equal to the stellar effective temperature $T_{\rm eff}$. Thus, we specify
\begin{equation}
   F_{\nu_0}=\frac{2\pi h\nu_0^3/c^2}{e^{h\nu_0/k_{\rm B}T}-1},
\end{equation}
and the frequency-integrated flux is given by 
\begin{equation}
    F = \sigma T^4,
\end{equation}
where $\sigma$ is the Stefan-Boltzmann constant. Since these are dependent on the wavelength of the transition, there is  a unique weighting parameter $\widetilde{W_i}$ for each line in every ion and temperature $T$. 

Using Equation (\ref{eq:W_i}), we can write the ratio of the bound to free acceleration as

\begin{equation}\label{eq:weight_func}
    \frac{g_{\rm bound} }{g_{\rm free}} = \frac{\chi_L}{\chi_e}\frac{\widetilde{W}_i}{\nu_0} = \frac{\chi_L}{\chi_e}\frac{\lambda_0\widetilde{W}_i}{c}.
\end{equation}

Finally, combining Equations (\ref{eq:chi_L}), (\ref{eq:Thomson opacity}), and (\ref{eq:weight_func}), we have
\begin{equation} \label{eq:gbound/gfree}
    \frac{g_{\rm bound}}{g_{\rm free}} = \frac{3}{8}\frac{\lambda_0}{r_e}f_{ij}\frac{n_i}{n_e}\left(1-e^{-h\nu_0/k_{\rm B}T}\right)\widetilde{W}_i,
\end{equation}
and we define the dimensionless line strength parameter $q_i$ as 
\begin{equation} \label{eq:q_i}
    q_i \equiv \frac{3}{8}\frac{\lambda_0}{r_e}f_{ij}\frac{n_i}{n_e}\left(1-e^{-h\nu_0/k_{\rm B}T}\right).
\end{equation}
This is similar to Equation 9 of \cite{Gayley1995}. However, here we have included the traditional correction factor for stimulated emission.

We also define the sum of the line strengths $\bar{Q}$ as
\begin{equation}
    \bar{Q} = \sum_i q_i \widetilde{W}_i.
\end{equation}

\subsection{Ionization Equilibrium}\label{subsec:Ion. Equilib.}

In order to calculate Equation (\ref{eq:q_i}), we first must first calculate the number density ratio $n_i/n_e$, which is given by 

\begin{equation}\label{eq:n_i/n_e}
\frac{n_{i}}{n_e} = \left(\frac{n_i}{n_{\rm ion}}\right)\left(\frac{n_{\rm ion}}{n_{\rm el}}\right)\left(\frac{n_{\rm el}}{n_H}\right)\left(\frac{n_H}{n_e}\right).
\end{equation}
In LTE, the total number of particles in the lower transition level can be described as a fraction of the total number of particles in the given ionization state:
\begin{equation}\label{eq:ni/nion}
    \frac{n_i}{n_{\rm ion}} = \frac{g_i e^{-(E_i - E_0)/k_{\rm B}T} }{U_{\rm ion}(T)}.
\end{equation}
Here $U_{ion}$ is the ion-specific temperature dependent partition function, $g_i$ is the ground state statistical weight, $E_i$ is the energy of the lower level of the transition, and $E_0$ is the ground level energy, which was set to zero by the atomic databases used here. It should also be noted that the oscillator strength $f_{ij}$ (as in Equation (\ref{eq:q_i})) is needed only to form the product $g_i f_{ij}$ (as in Equation (\ref{eq:ni/nion})) and never appears alone.

The elemental abundance ratio $n_{\rm el}/n_H$ was obtained from tabulated solar abundances \citep{Asplund2009},
whereas the other two quantities in parentheses, $n_{\rm ion}/n_{\rm el}$ and $n_H/n_e$, were found using the Saha equation:
\begin{equation} \label{eq:Saha}
    \frac{n_{i+1}}{n_i}n_e = \frac{2}{\lambda_e^3}\frac{U_{i+1}}{U_i}\exp{\left[-\frac{E_{i+1}-E_i}{k_{\rm B}T}\right]},
\end{equation}
where the thermal de Broglie wavelength of a free electron $\lambda_e$ is given by
\begin{equation}
    \lambda_e = h/ \sqrt{2\pi m_e k_{\rm B} T}.
\end{equation}

\begin{figure}
    \centering
    \includegraphics[width=\linewidth]{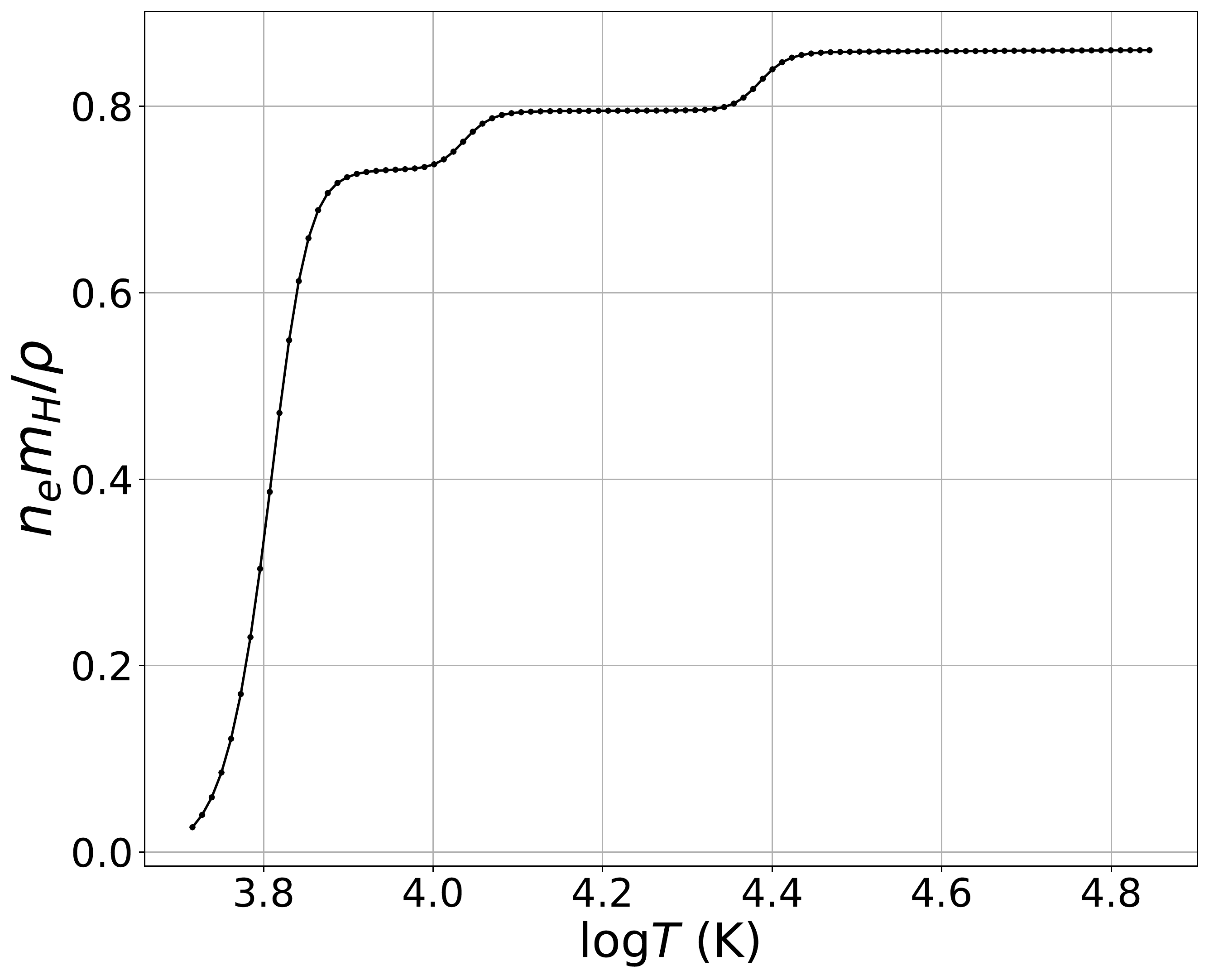}
    \caption{Iterated values of $n_e$ as a function of temperature, for $\rho = 10^{-13}$ g~cm$^{-3}$.}
    \label{fig:temp_v_ne}
\end{figure}

To solve Equation (\ref{eq:Saha}) for the pairwise ionization fractions $n_{i+1}/n_i$, we also need to know the electron number density $n_e$. Therefore, we made an initial estimate for $n_e$, which was then refined using an undercorrection technique. This was done at the end of each iteration over the Saha equation by tabulating a new estimate of $n_e$ from the calculated ionization balance, which was then multiplied by the previous estimate. The square root of this product was then used as the estimate of $n_e$ for the next iteration. We found that 20 iterations were sufficient to reach a stable value for $n_e$. An example set of calculations is shown in Figure \ref{fig:temp_v_ne}, which shows the final converged-upon values of $n_e$ for our temperature range, for an example density of $\rho=10^{-13}$ g~cm$^{-3}$.

The initial estimate for $n_e$ was given by 
\begin{equation}
    n_e = 0.1\left(\frac{\rho}{m_H}\right)
\end{equation}
where $\rho$ is the density of the wind and $m_H$ is the mass of Hydrogen. For this work, we used a grid of density with 22 values spanning $10^{-20}$ to $10^{-10}$ g~cm$^{-3}$, a similar density range to that used by \cite{Abbott1982}. This broad range is applicable to both massive star winds as well as other astrophysical environments that exhibit line-driven outflows. 

The quantity $n_{\rm el}$ can be found from the initial elemental abundances, here taken from \citet{Asplund2009}. These are given in the form of number density ratios to the abundance of hydrogen (i.e. $n_{\rm el}/n_{\rm H}$). From these we find the fractional abundance by mass $\mu$ of each element:

\begin{equation}\label{eq:n_el_1}
    \mu = \frac{A_{\rm el}\left(n_{\rm el}/n_{\rm H}\right)}{\sum_{\rm el} A_{\rm el}\left(n_{\rm el}/n_{\rm H}\right)}
\end{equation}
where $A_{\rm el}$ is the atomic weight of the element. The number density in cm$^{-3}$ for each element can then be found from the total mass density $\rho$, the fractional abundance $\mu$, and the mass of hydrogen $m_{\rm H}$:

\begin{equation}\label{eq:n_el_2}
    n_{\rm el} = \frac{\rho \mu}{A_{\rm el} m_{\rm H}}.
\end{equation}

\begin{figure}[htb!]
    \centering
    \includegraphics[width=\linewidth]{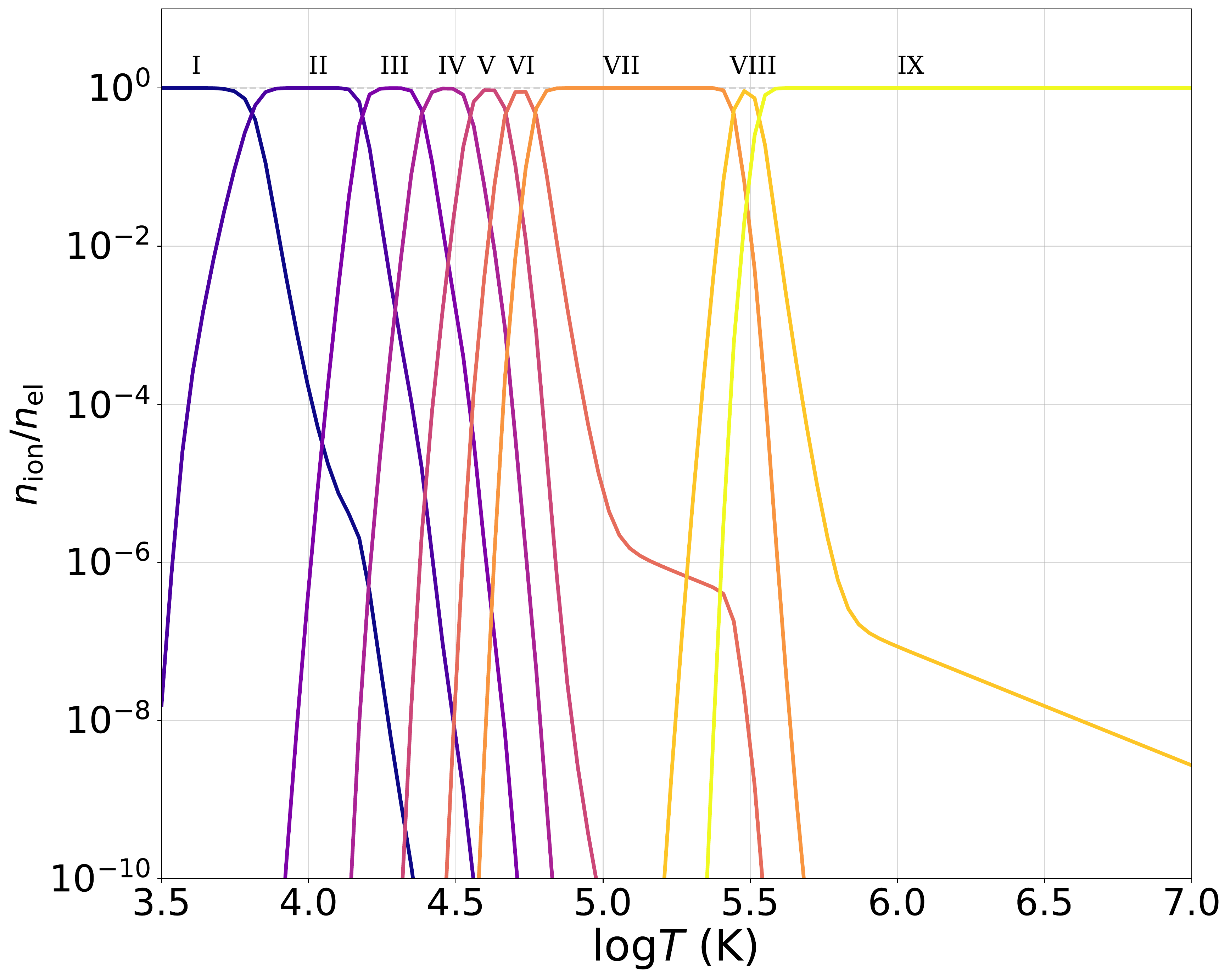}
    \caption{The ionization fraction $n_{\rm ion}/n_{\rm el}$ for the ionization stages of oxygen, for a fixed wind density of $\rho = 10^{-13}$ g~cm$^{-3}$.}
    \label{fig:ionization_frac_ox}
\end{figure}

Knowing $n_e$ and the tabulated number density of hydrogen $n_H$ from Equations (\ref{eq:n_el_1})--(\ref{eq:n_el_2}), we can then calculate the quantity $n_H/n_e$, leaving only the second quantity in Equation (\ref{eq:n_i/n_e}), $n_{\rm ion}/{n_{\rm el}}$. For ionization stage I this is given by 
\begin{equation}
  \frac{n_{\rm I}}{n_{\rm el}} = \left[\frac{n_{\rm II}}{n_{\rm I}}+\frac{n_{\rm III}}{n_{\rm II}}\frac{n_{\rm II}}{n_{\rm I}}+...\right]^{-1}  ,
\end{equation} 
where II and III represent ionization stages I and II of the element in question. Here $n_{\rm I}$ represents $n_{\rm ion}$ as seen in Equation (\ref{eq:n_i/n_e}). Each fraction in the brackets is given by the Saha equation (Equation (\ref{eq:Saha})).
We can similarly isolate $n_{\rm II}/n_{\rm el}$ and the higher ratios to find $n_{\rm ion}/n_{\rm el}$ for any ionization stage, and consequently $n_{\rm ion}/n_e$ and $n_i/n_e$ as in Equation (\ref{eq:n_i/n_e}), of any element that we consider. Figure \ref{fig:ionization_frac_ox} shows an example calculation of $n_{\rm ion}/n_{\rm el}$ for the ionization stages of oxygen. These steps were carried out for all elements from H to Ni. We considered all ionization stages for elements with atomic numbers $Z \leq 10$, and only the first ten ionization stages for elements with $Z > 10$.

\section{Atomic Data}\label{sec:Atomic Data}

\subsection{Partition Functions}\label{subsec:partition funcs}
The LTE partition functions used in Equations (\ref{eq:ni/nion})--(\ref{eq:Saha}) were calculated according to the fitting procedure set forth by \citet{Cardona2010}. Namely, for elements with $Z \leq 20$, we used the tabulated fit parameters in the expression
\begin{multline}
    U = g_{1jk} + G_{jk} e^{-\epsilon_{jk}/k_{\rm B} T} \\ +\frac{m}{3}(n_*^3-343)e^{-E_{n_*jk}/k_{\rm B} T}
\end{multline}
where $g_{1jk}$ is the ground state statistical weight, and $E_{n_*jk}$ is given by 
\begin{equation}
    E_{n_*jk} = \chi_{jk} - \frac{Z_{\rm eff}^2\mbox{Ry}}{n_*^2}
\end{equation}
for ionization state $j$ of element $k$. Here $\chi_{jk}$ is the ionization potential, Ry is the energy
of one Rydberg (13.6 eV), and $Z_{\rm eff}$ is the effective ion charge $j+1$. The effective maximum upper level index  $n_*$ is given by
\begin{equation}
    n_* = \frac{q}{2}\left(1 + \sqrt{1+\frac{4}{q}}\right), \textrm{ with } q = \sqrt{\frac{Z_{\rm eff}}{2\pi a_0}}n_{\rm tot}^{-1/6}
\end{equation}
where $a_0$ is the Bohr radius. The total number density of the gas $n_{\rm tot}$ is given by 
\begin{equation}
    n_{\rm tot} = n_e + \sum n_{\rm el},
\end{equation}
with $n_e$ given by the iterative undercorrection process described above, and $n_{\rm el}$ given by Equation (\ref{eq:n_el_2}). 

The quantities $m$, $G_{jk} \textrm{ and } \epsilon_{jk}$ are drawn from Table 1 of \cite{Cardona2010}. The values of $g_{1jk}$ and $\chi_{jk}$ are taken from CHIANTI for all modeled
elements and ionization stages (see Section \ref{sec:Databases}). \cite{Cardona2010} does not provide the fit parameters for the partition function for elements of $Z>20$, and a simple fitting procedure on the available parameters was performed to empirically calculate estimates of these parameters for higher-$Z$ elements. The Cardona $\epsilon_{jk}$ parameter correlated fairly well with ionization potential with no more than $\sim$20\% error, while the $m$ and $G$ parameters are well correlated with each other and weakly correlated with the ground state statistical weight $g_{1jk}$. These fits are given by
\begin{equation}
    \epsilon_{jk} = \chi_{jk} (0.946 - 0.007 Z_{\rm eff}),
\end{equation}
\begin{equation}
    m =  4 (g_{1jk})^{0.79},
\end{equation}
and
\begin{equation}
    G_{jk} = 113 m^{ 0.66}.
\end{equation}
These were used for all considered elements with $Z>20$. Because these new fitting parameters were developed for the first time here, we first investigated their impact by repeating the partition-function calculations with a simpler low-temperature approximation for elements with $Z > 20$: $U \approx g_{1jk}$. Doing this for the grids of temperature and density discussed below resulted in only negligible differences in the values of $U$.

\subsection{Database Selection}\label{sec:Databases}
To find the total radiative force on a parcel of gas for a given temperature and density, we need to construct the most complete line list of atomic data possible, as all spectral lines encountered by the radiation field must be accounted for. To this end, we compiled spectral line data from four sources: the National Institute of Standards and Technology (NIST) \citep{NIST_ASD}, version 9.0 of the CHIANTI database \citep{CHIANTI_1,CHIANTI_v9}, the database of lines used by the radiative transfer code CMFGEN \footnote{Atomic data used here are those which were updated by D.J. Hillier in 2016 (\url{http://kookaburra.phyast.pitt.edu/hillier/cmfgen_files/atomic_data_15nov16.tar.gz}).} \citep{Hillier1990,H&M1998,H&L2001}, and the Opacity Project's TOPbase \citep{TOPbase_1,TOPbase_2}. The use of multiple databases was necessary, as there exist gaps in the atomic data available from each individual database. 

We retrieved energy level classifications and wavelength data for each ion. For each selected element and ionization stage, tabulations of line oscillator strengths (i.e., $g_{i} f_{ij}$) were extracted, along with lower-level energies $E_i$ and rest-frame wavelengths $\lambda_0$. These are the parameters necessary to compute the line strength parameter $q_i$. We did this for all elements up to Ni. For the ionization stages, we retrieved data for each element up to nine times ionized (that is, data were retrieved for ionization stages I through X). For elements with atomic number $Z<10$, we retrieved data for all the available ionization states, up to fully ionized. In Appendix \ref{database appendix}, we summarize the process used to determine which database would be used for each ion. The final line list contains 4,514,900 lines. The detailed breakdown of line counts and the database used for each ion are given in Table \ref{Table: database by ion}.

\begin{figure}[htb!]
    \centering
    \includegraphics[width=\linewidth]{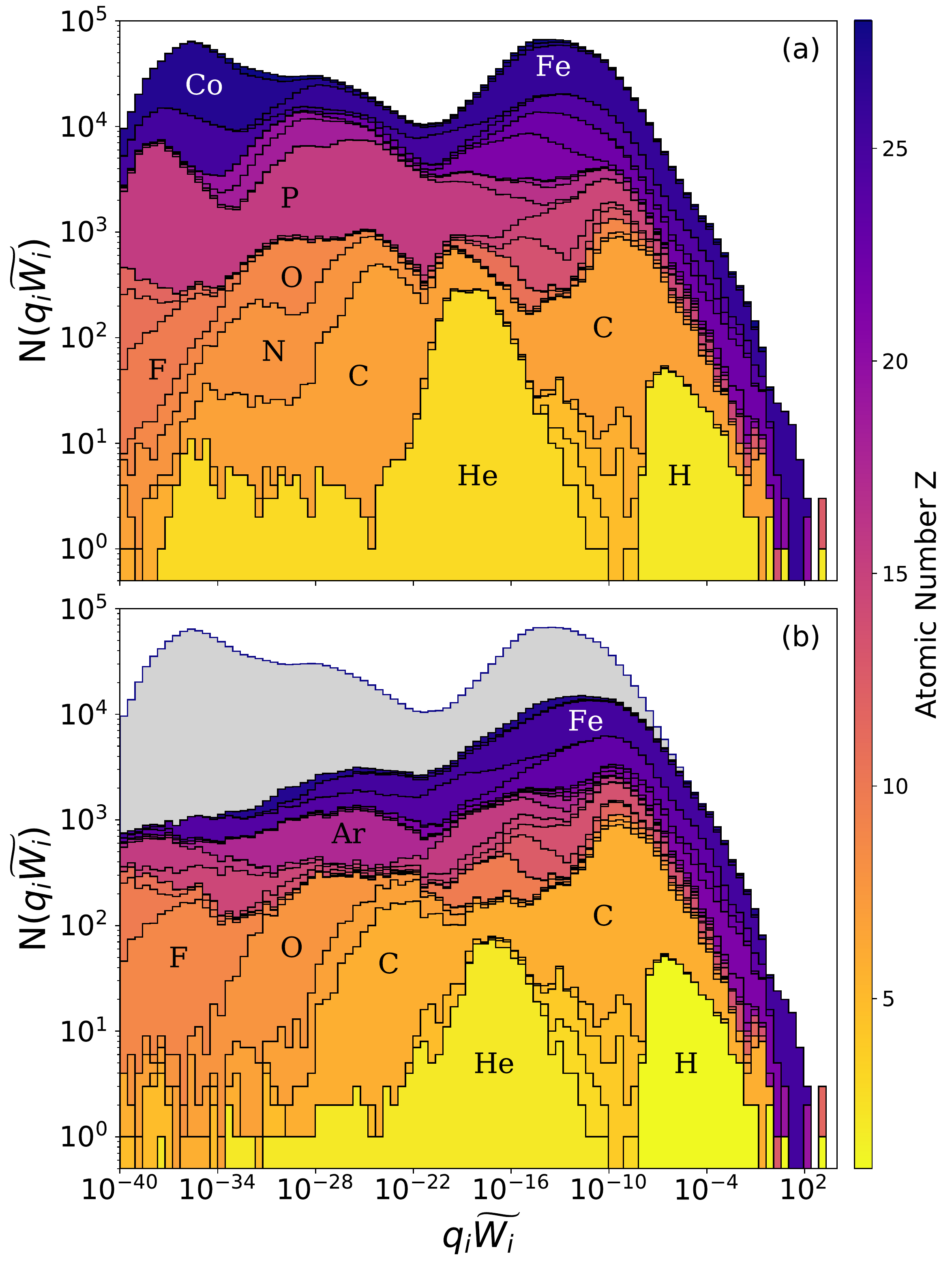}
    \caption{Comparison of the contributions by element to the total distribution of line strengths. (a) shows all transitions, including theoretical. (b) excludes theoretical transitions from the CMFGEN and CHIANTI databases. Major element contributions labeled, for example temperature $T=5200$~K and density $\rho = 10^{-13}$ g~cm$^{-3}$. In (b) grey indicates contour of the total histogram from (a).}
    \label{fig:sand plot obs comparison}
\end{figure}

Figure \ref{fig:sand plot obs comparison}(a) shows histograms for the occurrence frequency of different
values of $q_i \widetilde{W}_i$ for one example choice of the local temperature and density. All plotted histograms were constructed using 95 logarithmically spaced bins. Stacked underneath the uppermost curve (corresponding to the total counts in each bin) are individual histograms that break out the contributions from each individual element. In Figure \ref{fig:sand plot obs comparison}(b) we also show histograms computed from a subset of the CMFGEN and CHIANTI databases that includes only lines that have been observed experimentally (i.e., 589,186 lines out of the full set of 4,514,900 lines). Thus, this panel ignores the vast majority of lines (i.e., 87\%) with only theoretically predicted properties. There is a notable drop-off in line strengths below $q_i \widetilde{W}_i \approx 10^{-19}$ for the distribution that excludes the theoretical lines.
While the contributions by low-$Z$ elements are the same in both cases, this drop-off in line strengths represents a lack of observed lines for high-$Z$ elements, most notably cobalt. However, in both cases there is a significant contribution to the distribution by high-$Z$ elements, notably that of iron, at high line strengths. For the sake of completeness, we use the line list comprised of both theoretical and observed transitions for the remainder of this work. However, for the purpose of comparison Section \ref{sec:Force Multiplier} includes calculations of the force multiplier for both the full line list and the observed-only subset.

\section{Calculating and Fitting the Line Force Multiplier}\label{sec:Force Multiplier}
In order to examine the dynamics of the outflows, we first calculate the line force multiplier, a measure of the strength of the radiation force, from our updated list of spectral lines and distributions of $q_i\widetilde{W}_i$. We then fit and compare two functions to the resulting distributions.

\subsection{Calculating $M(t)$} \label{subsec:calcing M(t)}
The line acceleration is defined as the radiative acceleration due to electron scattering, multiplied by the line force multiplier $M(t)$. Here, $t$ is an optical depth parameter that is independent of the line strength. It is given by
\begin{equation}\label{eq: CAK_t}
    t = \kappa_e \rho v_{\rm th} \left|\frac{dv}{dr}\right|^{-1}
\end{equation}
for expanding atmospheres \citep[CAK]{Sobolev1957,Sobolev1960}.
In the case of a static atmosphere, $t$ is equivalent to the electron scattering optical depth, while in the expanding case $t$ is less than this depth.

In an expanding wind, we cannot simply take the sum of the $q_i\widetilde{W_i} = g_{\rm bound}/g_{\rm free}$ as found in Equation (\ref{eq:gbound/gfree}) above. The full force multiplier depends on other radiative transfer effects, such as the ``self-shadowing'' of the lines, due to differences in the Doppler-shifted local reference frames \citep{Gayley1995}. 
Therefore the full calculation of the force multiplier $M(t)$ from the line list can be written as 
\begin{equation}\label{eq: line list M(t)}
  M(t) \, = \, \eta \, \sum_{i} q_i \widetilde{W}_i
  \left( \frac{1 - e^{-\tau_i}}{\tau_i} \right)
\end{equation}
where the geometrical finite-disk factor $\eta$ is the ratio of the true line force to that derived in the limit of purely radial photons (e.g., it is the same as the ratio $F$ given by Equation (21) of \citealt{Gayley1995}). Equation (\ref{eq: line list M(t)}) assumes the Sobolev approximation and no overlapping lines, for supersonic flows. The dimensionless optical depth $\tau_i$ of each line is defined as
\begin{equation}\label{eq: tau_1}
    \tau_i = \frac{c}{v_{\rm th}}q_i t
\end{equation}
where $v_{\rm th}^2 = 2k_{\rm B}T / m_p$ is the proton thermal speed for all ions of a single temperature. 

$M(t)$ corresponds to the sum over the spectral lines that contribute to the wind. Originally, $M(t)$ was parameterized by \cite{Castor1974} in terms of the optical depth, depending only on the structure of the wind. However,  CAK
proposed that $M(t)$ could take the form of a power-law, expressed as 
\begin{equation}\label{eq:complete CAK M(t)}
  M(t) \, = \, \eta \, k t ^{-\alpha}
\end{equation}
where $k$ and $\alpha$ are the fit constants of the power-law. Because the finite disk factor $\eta$ appears in both Equations (\ref{eq: line list M(t)}) and (\ref{eq:complete CAK M(t)}), we can safely neglect it when performing fits and parameterizations for quantities such as $k$ and $\alpha$. We re-examine the assumption presented in Equation (\ref{eq:complete CAK M(t)}) that the line force multiplier takes the form of a power law. To do this, we will fit both power laws and an alternative fitting function to the values of $M(t)$ calculated from the updated line list. 

Although the CAK $t$ parameter can be evaluated at any point in a radiatively-driven outflow, here we  evaluate $M(t)$ over a logarithmic grid of $t$ spanning from $10^{-15}$ to $10$. In the limit of $t\rightarrow 0$
the line-force multiplier $M(t)$ should become equal to $\bar{Q}$. It should be noted that \cite{Abbott1982} considers only values of $t$ from $10^{-7}$ to $t=0.1$. However, we include values beyond this range in order to better examine the behavior of $M(t)$. For example, the asymptotic behavior of $M(t)$ in the limit of small $t$ is not seen in the range used by \citet{Abbott1982}. 

\begin{figure*}[htb!]
    \centering
    \includegraphics[width=\textwidth]{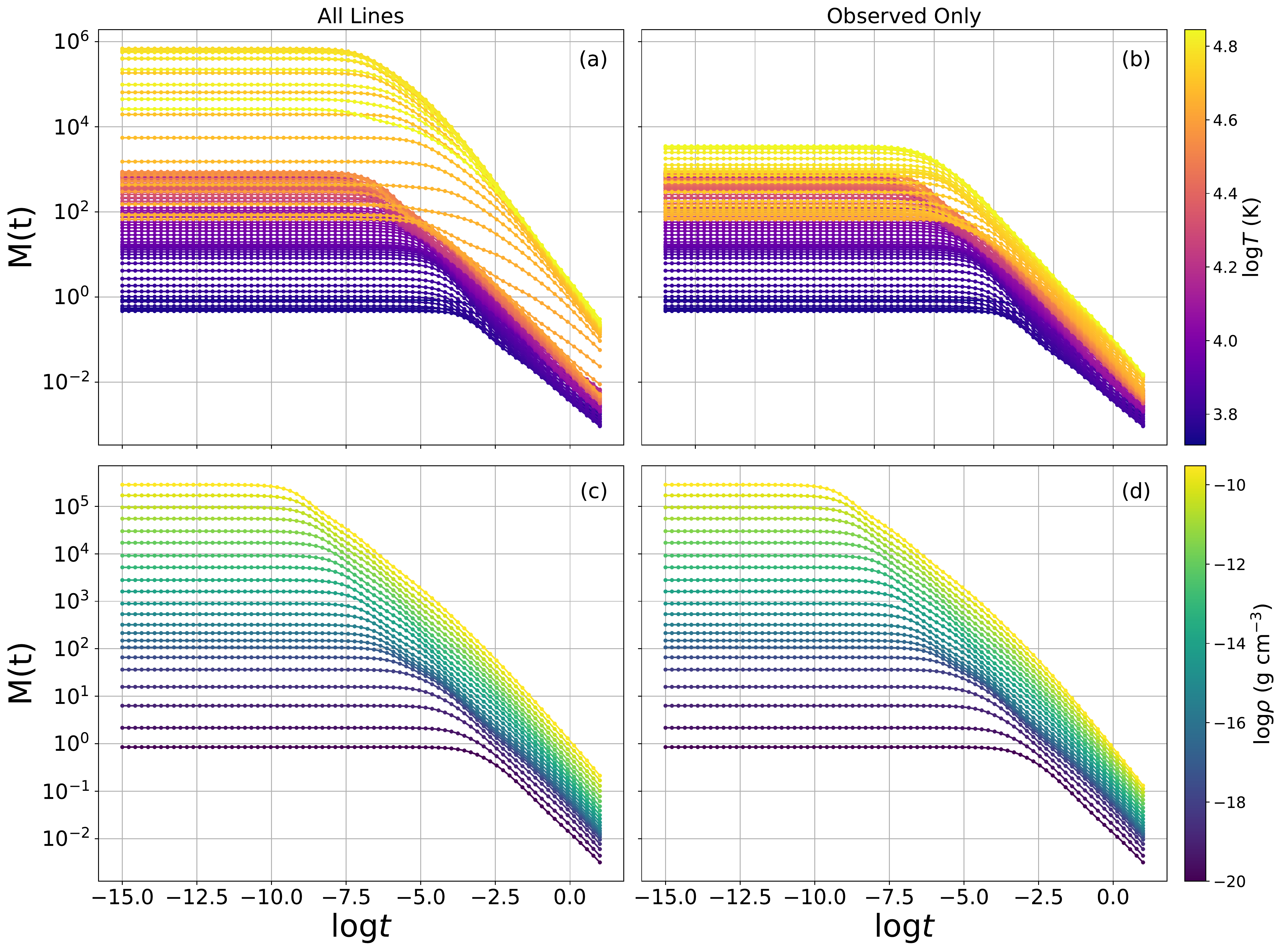}
    \caption{The force multiplier $M(t)$: (a) Varying with temperature from $T = $5,200~K to $T = $70,000~K for a constant density of $\rho=10^{-20}$ g~cm$^{-3}$, for observed and theoretical transitions. (b) Varying with temperature from $T = $5,200~K to $T = $70,000~K for a constant density of $\rho=10^{-20}$ g~cm$^{-3}$, for only observed transitions. (c) Varying with density from  $\rho = 10^{-20}$ g~cm$^{-3}$ to $\rho = 10^{-10}$ g~cm$^{-3}$ for a constant temperature of $T=5,200$~K, for theoretical and observed transitions. (d) Varying with density from  $\rho = 10^{-20}$ g~cm$^{-3}$ to $\rho = 10^{-10}$ g~cm$^{-3}$ for a constant temperature of $T=5,200$~K, for only observed transitions.}
    \label{fig: M(t) vary}

\end{figure*}

In calculating the dimensionless line strength parameter $q_i$, the weighting function $\widetilde{W_i}$, and subsequently the force multiplier $M(t)$ as described above, we used a grid of 100 logarithmically spaced temperatures ranging from 5,200 to 70,000 K and 22 densities ranging from $10^{-20}$ to $10^{-10}$ g~cm$^{-3}$. We calculated the force multiplier $M(t)$ for two cases: (1) the full
line list that includes both observed and theoretical lines, and (2) only
the subset of observed lines with laboratory wavelengths.
Figure \ref{fig: M(t) vary} shows results for $M(t)$ for both cases. In general, note that both higher temperatures and higher densities tend to result in higher values of the force multiplier. Note also that the turnover to a constant value of $M(t) \approx \bar{Q}$ occurs at different values of $t$; this is sometimes as low as $t \approx 10^{-10}$ and sometimes as high as $t \approx 10^{-4}$. \citet{Stevens&Kallman1990} calculated the radiative force due to X-ray ionization on the stellar wind in massive X-ray binaries. The flattening of the $M(t)$ curve seen in Figure \ref{fig: M(t) vary} is similar to Figure 1 of that paper, where the force multiplier is suppressed with increasing X-ray ionization.

\begin{figure}[htb!]
    \centering
    \includegraphics[width=\linewidth]{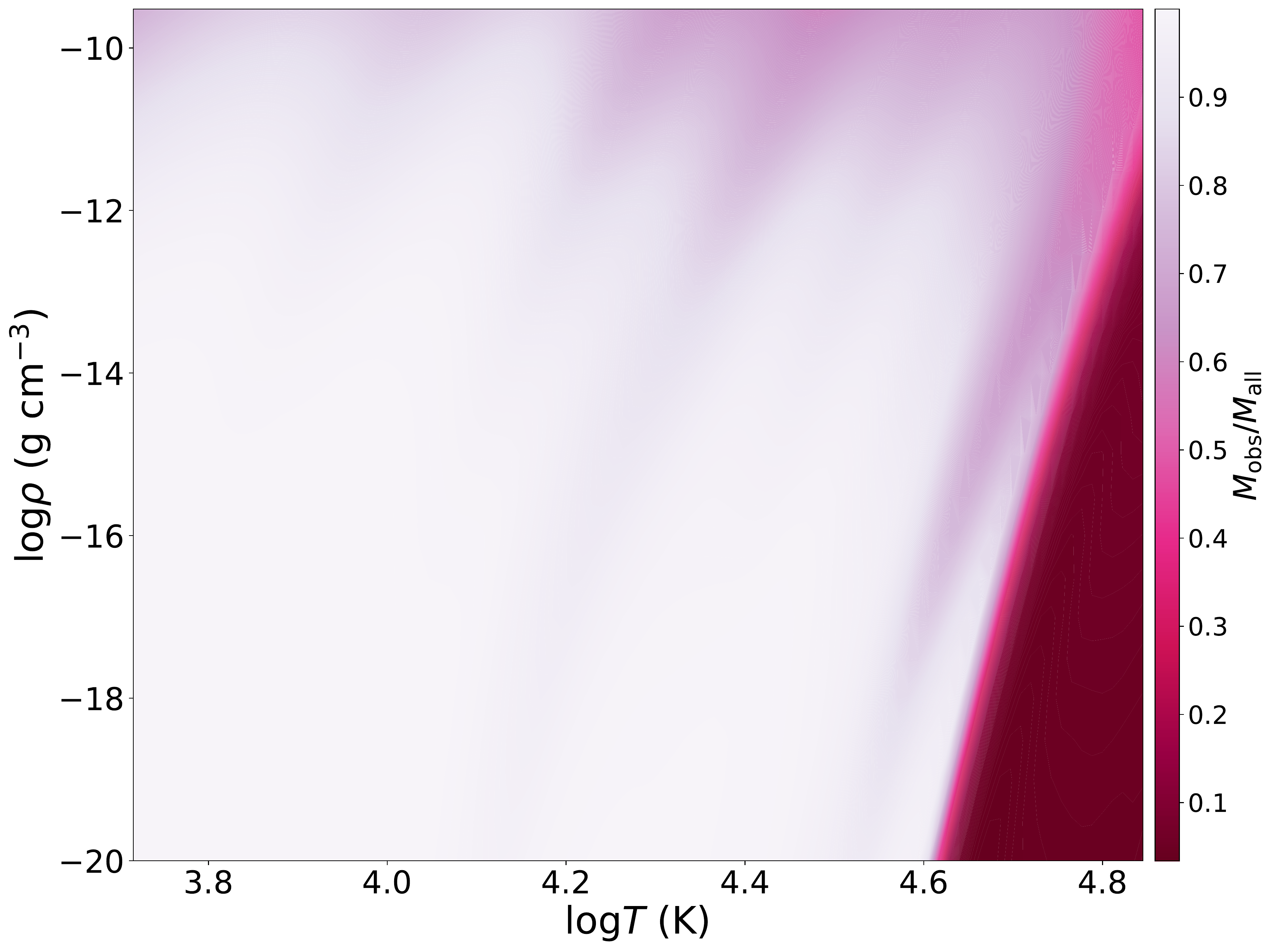}
    \caption{Ratio of the calculated force multiplier for line lists comprised of all lines to observed lines only, shown here for an example value of $t \approx 1$. }
    \label{fig: M_obs compare contour}

\end{figure}

Figure \ref{fig: M_obs compare contour} shows the ratio of the full and observed-only calculations for an example value of $t\approx 1$. The discrepancy between the two multipliers is most pronounced at high temperatures, with the full line list producing a much higher force multiplier than the observed-only list. This is to be expected, as the observed-only list does not include many lines from high-$Z$ elements which require high temperatures to ionize. However, at lower temperatures the discrepancy between the two calculations is slight. In light of these factors, going forward we use only the full line list that includes theoretical lines. 

\begin{figure}[h!]
    \centering
    \includegraphics[width=\linewidth]{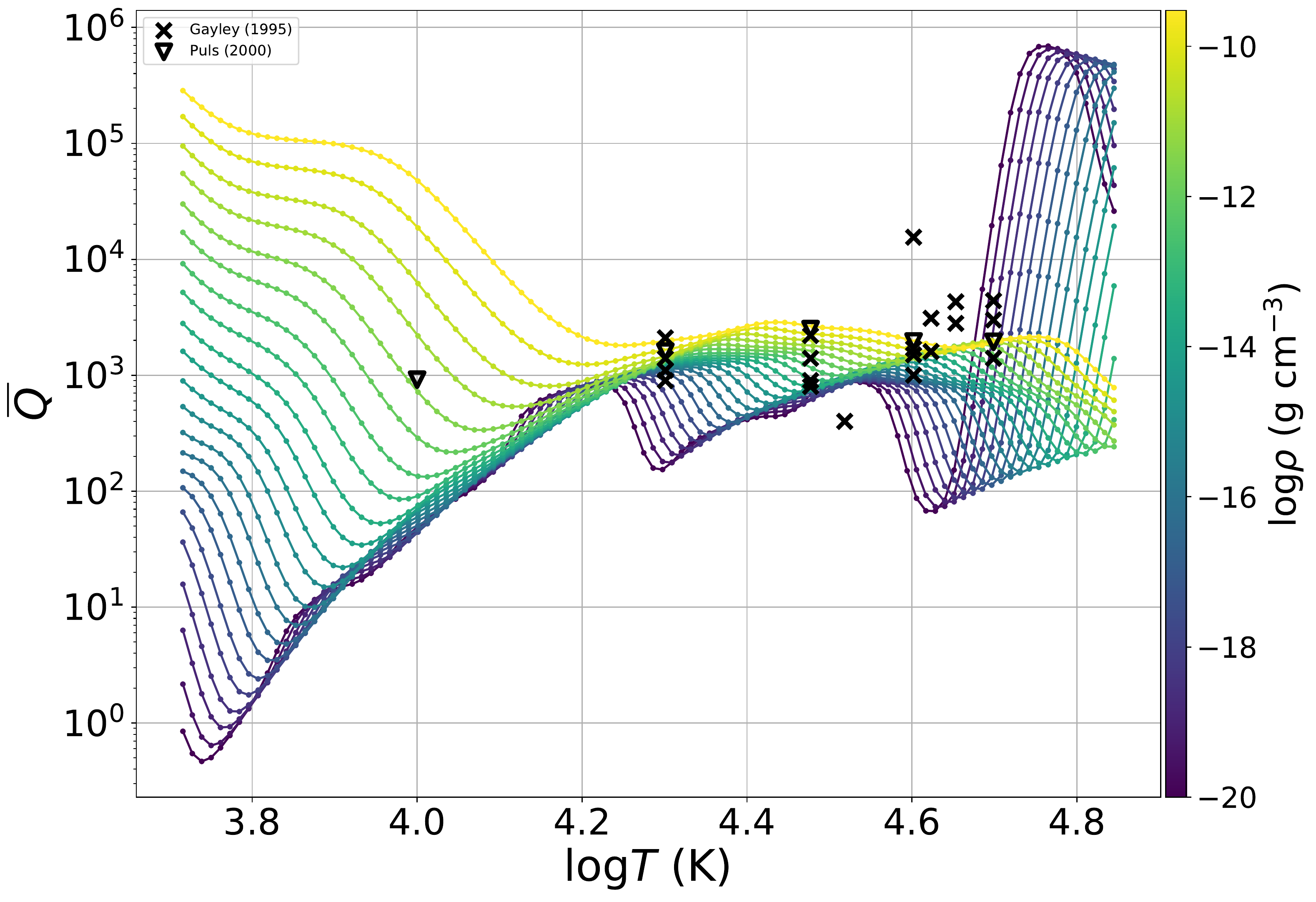}
    \caption{Evolution of $\bar{Q}$ for our chosen density and temperature range, compared to values from Table 1 of \cite{Gayley1995} (black crosses) and Table 2 of \cite{Puls2000} (black triangles).}
    \label{fig: Qbar}
\end{figure}

Figure \ref{fig: Qbar} shows the sum $\Bar{Q} = \sum_i q_i\widetilde{W_i}$, which was calculated for each temperature and density. These are compared to values from \cite{Gayley1995}, who also provided values of $\bar{Q}$ as converted from previous works such as \cite{Abbott1982}. Note that temperatures corresponding to O and early-B spectral types ($\log T$ between 4.2 and 4.6) tend to exhibit values of $\bar{Q}$ around $10^3$, independent of density, as also found by \citet{Gayley1995}. Above and below this range, there is a strong dependence of $\bar{Q}$ on density, with $\bar{Q}$ varying drastically at both very low temperatures ($\log T \lesssim 4.0 $), and very high temperatures ($\log T \gtrsim 4.7$). This probably indicates that departures from LTE Saha ionization balance are more important to take into account for these temperatures.

\subsection{CAK Power Law Fitting}\label{subsec: CAK fits}
We begin by fitting a power law of the form proposed by CAK (Equation (\ref{eq:complete CAK M(t)})) to the calculated values of $M(t)$. This fitting was performed using the Levenberg-Marquardt method of least-squares fitting. As is readily apparent from Figure \ref{fig: M(t) vary}, the full range of calculated $M(t)$ values cannot be described by a single power law. Therefore, we fit the initial CAK power law form to only to values of $\log(t)>-3$, in order to exclude the flat portion of the curve.

\begin{figure}[hbt!]
    \centering
    \includegraphics[width=\linewidth]{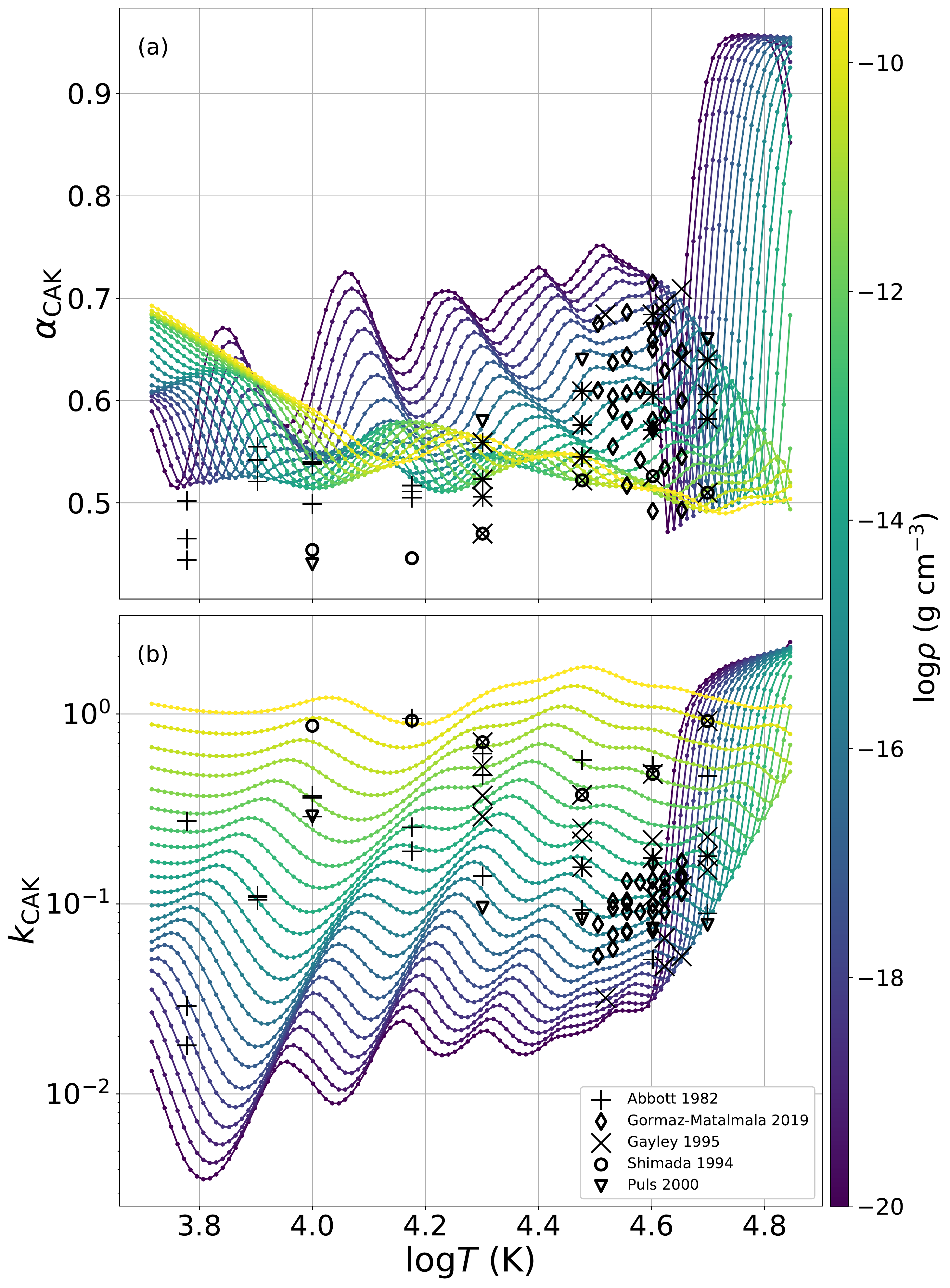}
    \caption{Comparison of CAK power law parameters $\alpha$ (a) and $k$ (b) with values from \cite{Abbott1982}, \cite{Shimada1994}, \cite{Gayley1995}, \cite{Puls2000}, and \cite{Gormaz-Matamala2019}, for the our chosen range of densities.}
    \label{fig: alpha and k comp with lit}
\end{figure}

As with $\bar{Q}$, we also compare the fitted values of $\alpha$ and $k$ to those from previous work. Figure \ref{fig: alpha and k comp with lit} shows a comparison to \cite{Abbott1982}, \cite{Shimada1994}, \cite{Gayley1995}, and \cite{Gormaz-Matamala2019}, with a good agreement with those values, most especially for densities that fall in the middle of our considered density range. Note that for the lowest values of $T$, our values for the power-law slope $\alpha$ never get as small as some of the plotted literature values that are of order $\sim$0.45. This may be explained if the power-law fits performed in other papers included portions of the flattened parts of $M(t)$. For example, \cite{Abbott1982} calculated values of $\alpha$ and $k$ for a range of $-6 \leq \log(t) \geq -1$ for a density grid similar to our own. We find that for low temperatures, the flat portions of the $M(t)$ curve begin at values of $t$ as high as $\log(t)\sim -2.5$. Inclusion of these flat portions during fitting would yield shallower slopes than those found in this work.

Despite acceptable agreement with previous works for these two parameters ($\alpha$, $k$), this preliminary power law form of the force multiplier presents a decent fit to the calculated values of $M(t)$ for only a narrow range of $t$, namely between $\log(t) \approx -3$ and $\log(t) \approx 0$.

\subsection{Alternate Fitting Function} \label{subsec: alt fit function}
\begin{figure}[hbt!]
    \centering
    \includegraphics[width=\linewidth]{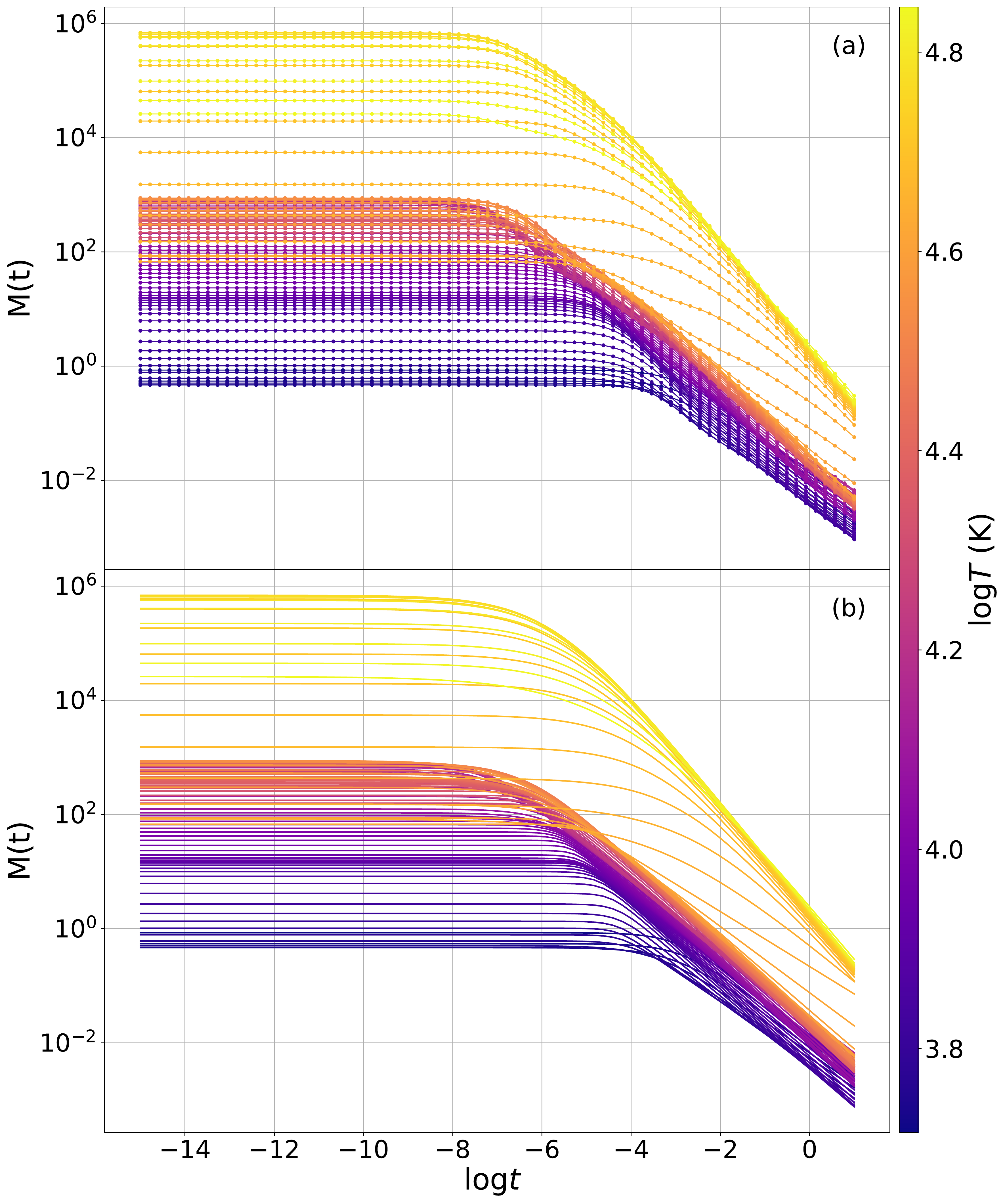}
    \caption{Comparison of alternate fits and calculated values of M(t) for an example density of $\rho = 10^{-20}$ g~cm$^{-3}$. (a) $M(t)$ as calculated from line list. (b) Fits produced by Equation (\ref{eq: Alt M(t)}).}
    \label{fig: Alt M(t) fits}
\end{figure}

We present an alternative fitting function in the form of a saturated power law, given by 
\begin{equation}\label{eq: Alt M(t)}
    M(t) = \frac{\eta \bar{Q}k}{(k^s + \bar{Q}^st^{\alpha s})^{1/s}}.
\end{equation}
In this case, $\alpha$, $k$, and $s$ are fit parameters, and $\bar{Q}$ is the calculated value as found from the line list and shown in Figure \ref{fig: Qbar}. The parameter $s$ is a sharpness parameter that determines how rapidly the function transitions from the low-$t$ to high-$t$ limits. This function reduces to the CAK power law form in the limit of large $t$, and in the limit of small $t$ reduces to $\bar{Q}$, consistent with the behavior of the calculated values of $M(t)$. Figure \ref{fig: Alt M(t) fits} shows the resulting fits compared to the calculated $M(t)$ values, for an example density of $\rho = 10^{-20}$ g~cm$^{-3}$. 
The above function was found to be flexible enough to fit the calculated $M(t)$ values quite accurately. For the full grid of parameters ($T$, $\rho$, and $t$) we computed the
fractional difference $D$ between the numerical and best-fit values of $M(t)$. The median of this distribution was $\langle D \rangle = 1.95$\%, and only a tiny fraction of the model values (1.03\%) had values of $D > 30$\%.

The fits of the alternative function and the CAK power law are comparable for values of $\log t \gtrsim -4$.  However, for values of $\log t \lesssim -4$, the fit achieved using the alternate function is drastically improved over that of the CAK power law. As can be seen in Figure \ref{fig: M(t) vary}, the value of $t$ at which the CAK power law begins to fail also strongly depends on temperature and density. Additionally, we do not include a high opacity cut-off in our new form of $M(t)$, such as that suggested by \cite{Gayley1995}. Upon calculation of the force multiplier out to high $t$ ($t\sim 10^{15}$), we find that $M(t)$ continues to decrease as a power-law rather than as an exponential drop-off at high opacities, generally approaching the form $M(t) \propto t^{-1}$ as temperature and density increase. Therefore, we do not impose a cut-off of the force multiplier at high opacities.

\begin{figure*}[hbt!]
    \centering
    \includegraphics[width=\textwidth]{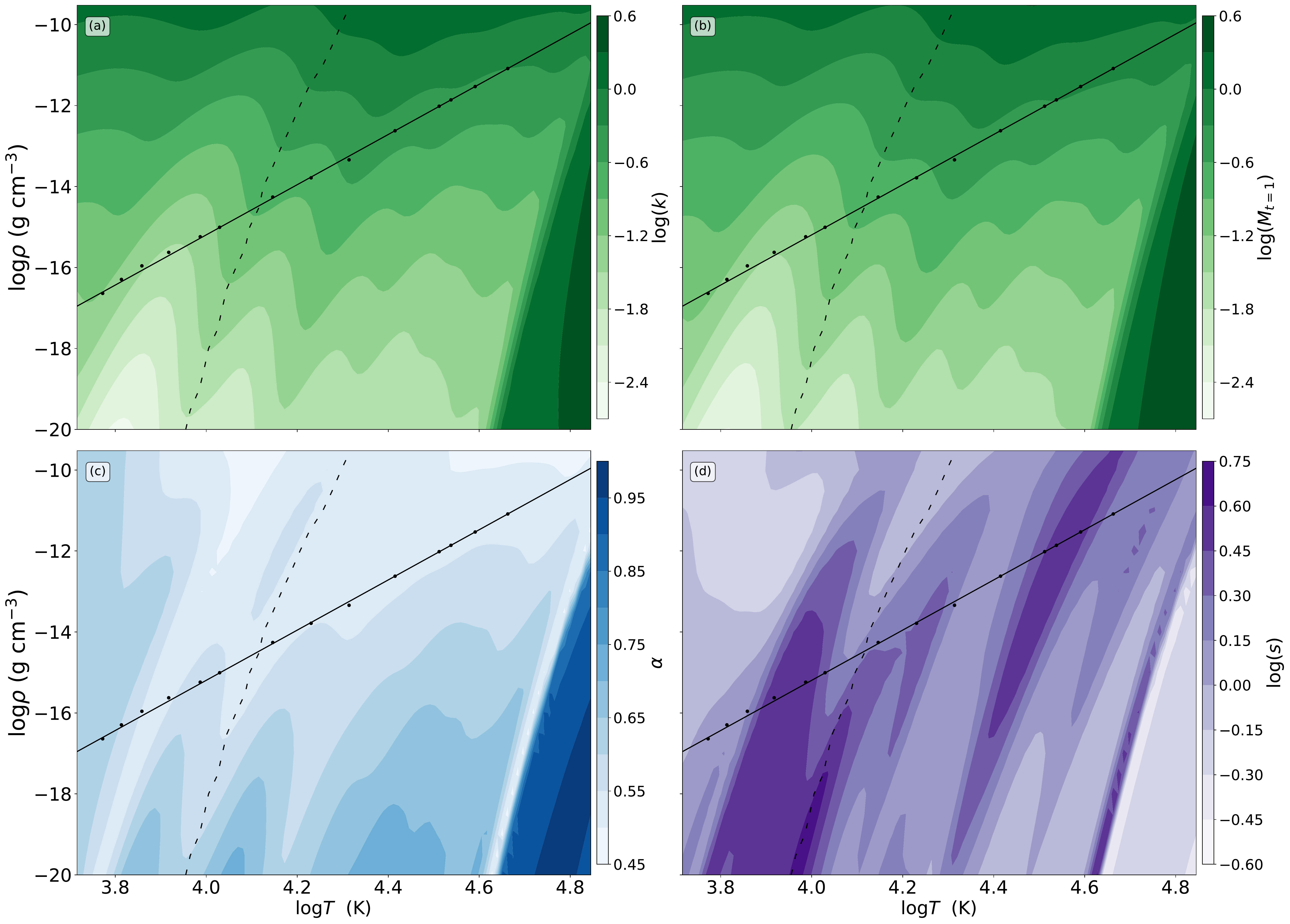}
    \caption{Two-dimensional contour plots of the temperature and density dependence of the fit parameters for the alternate fitting function. (a) $k$, (b) $M(t=1)$, (c) $\alpha$, (d) $s$. Solid lines indicate estimates of the density at the CAK critical point (Equation (\ref{eq: rho_crit})), and dashed lines indicate the calculated recombination temperature of Fe between Fe~III and Fe~IV.}
    \label{fig: Alt parameter evolution}
\end{figure*}

Figure \ref{fig: Alt parameter evolution} shows the dependence of the saturated power law fit parameters $\alpha$, $k$, and $s$
on temperature and density. Also shown is the evolution with temperature and density of $M(t)$ at $t=1$. At high values of $t$ such as $t=1$, the force multiplier behaves as a power-law. The values of $M(t=1)$ provide an alternate estimation of the fit parameter $k$. Comparison of Figures \ref{fig: Alt parameter evolution}(a) and \ref{fig: Alt parameter evolution}(b) additionally provide insight into the deviation of the fitted saturated power-law (a) from the actual calculated value (b). We also include an estimate of the CAK critical point density across the range of effective temperatures (see Section \ref{subsec: critical point}). Additionally, we indicate a locus of parameters at which the Saha ionization balance produces equal amounts of Fe~III and Fe~IV. The relevance of this to a proposed explanation for the phenomenon of ``bistability'' in line-driven winds is discussed further in Section 6.

\section{Mass-Loss Estimates for Massive Stars}\label{sec: Mass-Loss Rates}
In addition to determining a function to better describe the line force multiplier $M(t)$, we also explore the consequences of this updated form on the calculated mass-loss rates of massive stars. We do this by calculating mass-loss rates for both the traditional power-law form and our newly updated alternate form.

\subsection{General Mass-Loss Solution}\label{subsec: general mass loss}

We begin with the time-steady radial component of the momentum conservation equation. To approximate the supersonic winds of OB stars, we omit the gas-pressure gradient term. Because gas pressure does not play a fundamental role in highly supersonic winds such as the ones considered in this work, we can safely neglect these terms \citep[see, e.g.,][]{Gayley2000,Owocki2004}. Doing this, we have
\begin{equation}\label{eq: momentum conservation}
    v\pderiv{v}{r} = -\frac{GM_*}{r^2} + g_{\rm free} + g_{\rm bound}.
\end{equation}
Here, $v(r)$ is the radially dependent wind velocity, $M_*$ is the mass of the central star, and $G$ is the gravitational constant. The free radiative acceleration due to Thomson scattering $g_{\rm free}$ can be written as the Eddington factor $\Gamma$ times the gravitational acceleration. With stellar bolometric luminosity $L_*$, the Eddington factor $\Gamma$ can be written as
\begin{equation}\label{eq: Gamma}
    \Gamma = \frac{\kappa_e L_*}{4\pi c G M_*}
\end{equation}
with the mixture-dependent Thomson scattering coefficient given by 
\begin{equation}\label{eq: kappe_e}
  \kappa_e \, = \, \frac{\sigma_T n_e}{\rho}
  \, \approx \, \frac{\sigma_T}{m_H} \left( \frac{1+X}{2} \right)
\end{equation}
where $X$ is the hydrogen mass fraction. The final approximation above is provided only for reference in the limit of full ionization \cite[see, e.g.,][]{Mihalas1978}. In all results shown below, we use the self-consistent values of $n_e$ computed from the Saha equation. For the values of $X$ and $Y$, we use the bulk composition chemical abundances of H and He given in Table 4 of \cite{Asplund2009}, with $X=0.7154$ and $Y=0.2703$. Using Equations (\ref{eq: momentum conservation})--(\ref{eq: kappe_e}), we can write the equation of motion as 
\begin{equation}\label{eq: eq of motion}
    v\pderiv{v}{r} = \frac{GM_*}{r^2}\left[-1 + \Gamma + \Gamma M(t)\right].
\end{equation}

As in \cite{Gayley1995}, we also define the dimensionless wind acceleration factor $w$:
\begin{equation}\label{eq:w}
    w = \frac{r^2 v}{GM_*(1-\Gamma)} \deriv{v}{r}.
\end{equation}
This allows us to write Equation (\ref{eq: eq of motion}) as 
\begin{equation}\label{F1}
    F_1 = w + 1 -\frac{\Gamma}{1-\Gamma}M(t) = 0.
\end{equation}
We use this form because the CAK  critical-point solution requires at least two conditions to be true for a time-steady wind:
\begin{equation}\label{eq: conditions}
    F_1 = 0  \textrm{ and } F_2 = \pderiv{F_1}{w} = 0.
\end{equation}
Using mass conservation, we can write the density as $\rho =\Dot{M}/(4\pi r^2 v)$. Combining the definitions of the CAK $t$ parameter and the wind acceleration factor $w$, as given in Equations (\ref{eq: CAK_t}) and (\ref{eq:w}) respectively, we can then write $t$ as $t = t_m / w$, where
\begin{equation}\label{eq: alt t def}
    t_m = \frac{v_{\rm th} c \Dot{M} \Gamma}{L_* (1-\Gamma)}.
\end{equation}
This is equivalent to Equation (52) of \cite{Gayley1995}. We now can solve Equation (\ref{eq: conditions}) for $w$ and $t_m$ to determine the mass-loss rate. 

\subsection{Mass-Loss Rates for the CAK Multiplier}\label{subsec: CAK mass losses}
We begin with the traditional CAK power-law form of the force multiplier as given in Equation \ref{eq:complete CAK M(t)}, assuming that the parameters $\alpha$ and $k$ are known for a given set of lines. Extending beyond CAK, we also include the finite disk factor $\eta$, which has a simple form for a CAK-like force multiplier when evaluated at the stellar surface:
\begin{equation}\label{eq: eta disk factor}
    \eta \approx \frac{1}{1+\alpha}
\end{equation}
\citep{Kudritzki1989}. Therefore we can write the critical point conditions $F_1$  and $F_2$ as
\begin{equation}\label{eq: F1 CAK}
    F_1 = w + 1 - Cw^\alpha = 0
\end{equation}
\begin{equation}\label{eq: F2 CAK}
    F_2 = 1 - \alpha Cw^{\alpha-1} = 0
\end{equation}
where $C = \eta \Gamma k(1-\Gamma)^{-1}t_m^{-\alpha}$. Solving $F_2$ for $C$ and re-solving $F_1$ for $w$, we find the analytic solutions
\begin{equation}\label{eq: CAK w}
    w = \frac{\alpha}{1-\alpha} \textrm{ and } C = \frac{1}{\alpha^\alpha(1-\alpha)^{1-\alpha}}.
\end{equation}
Using this solution for $C$, we can solve for $t_m$, which can then be converted to $\Dot{M}_{\rm CAK}$. 
Combining equations (\ref{eq: F1 CAK})--(\ref{eq: CAK w}) and recalling the definition of $t_m$ from Equation (\ref{eq: CAK_t}), we find the mass-loss rate $\Dot{M}_{\rm CAK}$:
\begin{equation}
    \Dot{M}_{\rm CAK} = \frac{L_*(1-\Gamma)}{v_{\rm th}c\Gamma}\left[\frac{\alpha^\alpha \eta\Gamma k (1-\alpha)^{1-\alpha}}{(1-\Gamma)}\right]^{1/\alpha}.
\end{equation}
It is worth mentioning here that the apparent dependence of $\dot{M}_{CAK}$ on the Doppler thermal width $v_{\rm th}$ is in fact only a fiducial dependence, due to the definition of the $t$ parameter (Equation (\ref{eq: CAK_t})), which introduces $v_{\rm th}$ and is subsequently present in Equation (\ref{eq: alt t def}). While a reformulation of $t$ could eliminate this dependence, we choose to carry it through our calculations in order to maintain a level of comparability with previous works, notably that of CAK.

\subsection{Mass-Loss Rates for Updated Formalism}\label{subsec: alt mass losses}
 Next, we solve the critical-point equations for our more general form of $M(t)$, given by Equation (\ref{eq: Alt M(t)}). 
Combining Equations (\ref{F1}) and (\ref{eq: conditions}), we find $F_1$ and $F_2$ are now given by
\begin{equation}\label{eq: Alt F1}
    F_1 = w+1 -B \left[\frac{kw^\alpha}{(k^s w^{\alpha s} + \bar{Q}^st_m^{\alpha s})^\frac{1}{s}}\right] = 0 
\end{equation}
\begin{equation}\label{eq: Alt F2}
    F_2 = 1-B\alpha \left[ \frac{(kw^{\alpha-1})(\bar{Q}^st_m^{\alpha s})}{(k^sw^{\alpha s}+ \bar{Q}^s t_m^{\alpha s})^{\frac{1}{s} +1 }} \right] = 0
\end{equation}
where $B = \eta \Gamma \bar{Q} / (1-\Gamma)$. Solving Equation (\ref{eq: Alt F1}) for
\begin{equation}\label{eq: Qt^salpha}
    \bar{Q}^st_m^{\alpha s} = \left(\frac{kw^\alpha B}{w+1}\right)^s - k^sw^{\alpha s}, 
\end{equation}
we substitute the result into Equation (\ref{eq: Alt F2}). This gives
\begin{equation}\label{eq: Alt quadratic}
    0 = wB^s + \alpha(w+1)\left[(w+1)^s -B^s\right].
\end{equation}
In the $\bar{Q}\rightarrow \infty$ ($B \gg 1$) limit, Equation (\ref{eq: Alt quadratic}) reduces to the CAK behavior, with a solution of $w = \alpha/(1-\alpha)$ as in Equation (\ref{eq: CAK w}). In the opposite limit ($B \rightarrow 0$), the above equation is solved
only by $w \approx -1$, which is unphysical (see below).

For the nominal case of $s = 1$, Equation (\ref{eq: Alt quadratic}) reduces to a quadratic equation with two unique solutions for $w$:
\begin{equation}\label{eq: s=1 quadratic}
    w = -\frac{B(1-\alpha)}{2\alpha}-1 \pm \frac{B(1-\alpha)}{2\alpha}\sqrt{1+\frac{4\alpha}{B(1-\alpha)^2}}
\end{equation}
However, in the more general case of $s\neq 1$ Equation (\ref{eq: Alt quadratic}) must be solved numerically. In this work this is done using the Newton-Raphson method. Once the acceleration factor $w$ is known, we can then find $t_m$ from Equation (\ref{eq: Qt^salpha}), which then in turn allows us to find the mass-loss rate $\dot{M}_{\rm alt}$ from Equation (\ref{eq: alt t def}).

It is relevant to note that in calculating the mass-loss rates that result from both the CAK and alternate forms, we discard any sets of parameters that result in $\Gamma \geq 1$ (since we do not include multiple-scattering effects or super-Eddington flows). Similarly, we discard as unphysical any negative solutions of the wind acceleration factor $w$. In Equations (\ref{eq: Alt quadratic}) and (\ref{eq: s=1 quadratic}) above, the condition $w \geq 0$ corresponds to $B \geq 1$. Since both $\eta$ and $(1-\Gamma)$ tend to be order-unity quantities, the condition for physically realistic solutions is thus $\Gamma \gtrsim \bar{Q}^{-1}$.
With typical values of $\bar{Q}$ of a few thousand, this implies that whenever $\Gamma$ drops below $\sim 10^{-3}$, a steady-state line-driven wind may not be possible.

\subsection{CAK Critical Point and Stellar Parameters}\label{subsec: critical point}
\begin{figure*}[hbt!]
    \centering
    \includegraphics[width=\textwidth]{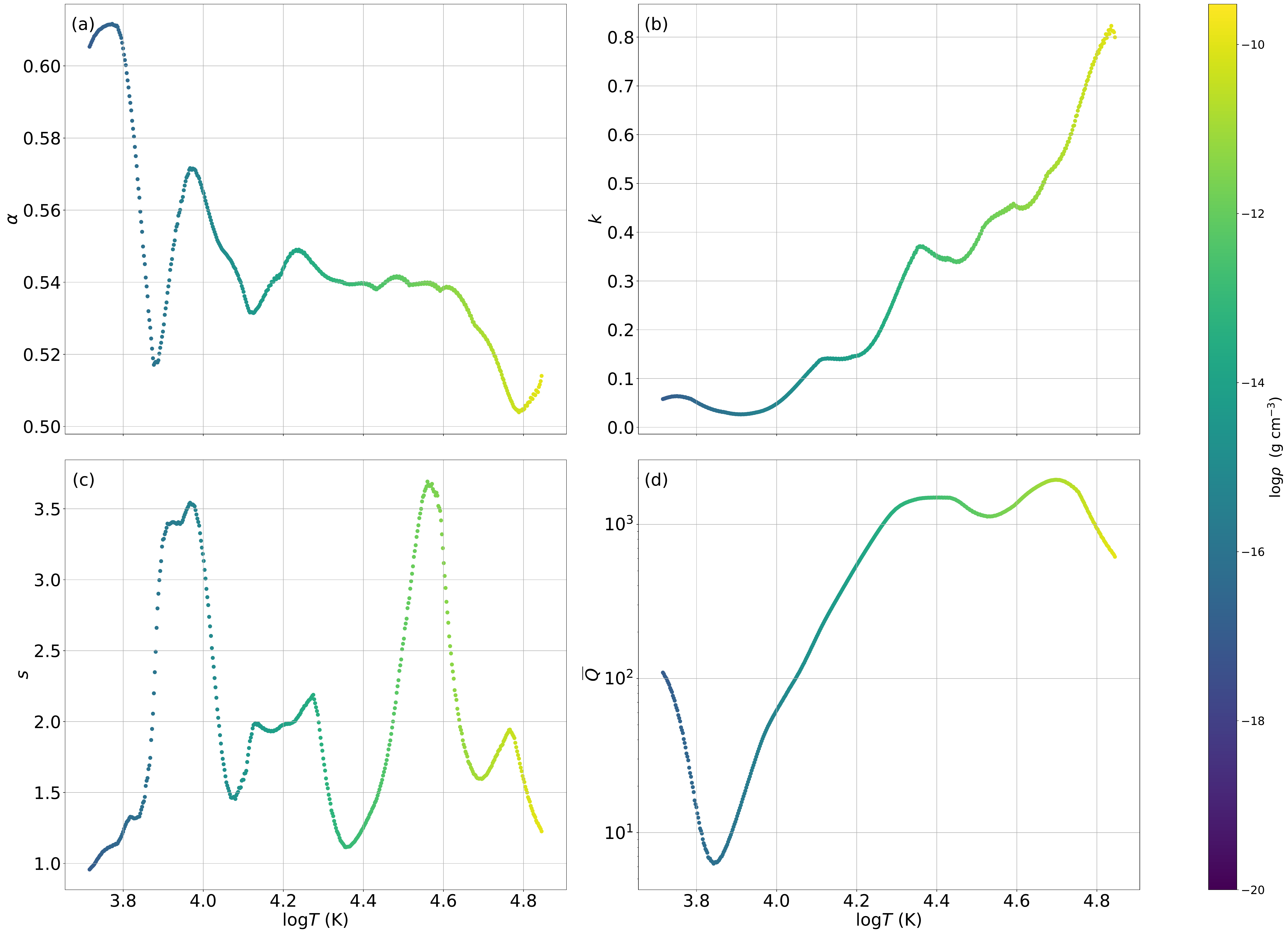}
    \caption{Evolution with temperature of the four parameters that define the saturated power-law form of $M(t)$ along the CAK critical point line. (a) $\alpha$, (b) $k$, (c) $s$, (d) $\bar{Q}$.}
    \label{fig: critical point parameter evol}
\end{figure*}

Although the force multiplier $M(t)$ is a function of both $T$ and $\rho$, it is possible to characterize much of the physics by evaluating $M(t)$ at the critical point of the flow \citep[see, e.g., CAK;][]{Abbott1980,Pauldrach1986,Bjorkman1995}. To do this, we need to know both the temperature and the density at the critical point. For an isothermal wind, $T$ at the critical point is given more or less by the photospheric effective temperature. The only way to estimate the density $\rho_{\rm crit}$ at the critical point, though, is to have an associated ``initial guess'' for the full radial dependence of the plasma parameters. We provide this initial guess for a set of idealized main-sequence stellar properties (see Section \ref{subsec: mass loss comparison}) by using a modified-CAK (mCAK) numerical code developed by \citet{Cranmer&Owocki1995}. This code solves the equations of mass and momentum conservation for the power-law CAK force multiplier and a standard version of the uniformly illuminated finite-disk factor $\eta$. When considering finite sound-speed effects, it is necessary to solve simultaneous singularity and regularity conditions for the properties of the critical point (CAK).

For a sequence of stellar properties spanning effective temperatures between 5,920 and 46,000~K (see below for details), we produced mCAK models with fixed line-force constants $\alpha = 0.6$ and $k = 0.5$. These models all exhibited critical points at radial distances between 1.01 and 1.02 times the photospheric stellar radius, critical wind speeds between about 50 and 120 km~s$^{-1}$ (i.e., typically about 3\% of the asymptotic or terminal wind speeds of 2,000--3,000 km~s$^{-1}$), and values of $\rho_{\rm crit}$ between $10^{-17}$ and $10^{-11}$ g~cm$^{-3}$. Figure \ref{fig: Alt parameter evolution} shows this trend in two-dimensional ($T$, $\rho$) diagrams. A power law of the form 
\begin{equation} \label{eq: rho_crit}
    \rho_{\rm crit} = \left(6.33\times 10^{-16}\textrm{ g~cm}^{-3}\right)\left(\frac{T}{10^4 \textrm{ K}}\right)^{6.2}
\end{equation}
is reasonably successful at capturing this trend as well. Figure \ref{fig: critical point parameter evol} shows how the calculated parameter $\bar{Q}$ and the fit parameters $\alpha$, $k$, and $s$ vary with temperature along the CAK critical point curve, which corresponds to the black dashed line shown in Figure \ref{fig: Alt parameter evolution}.

\begin{figure}[hbt!]
    \centering
    \includegraphics[width=\linewidth]{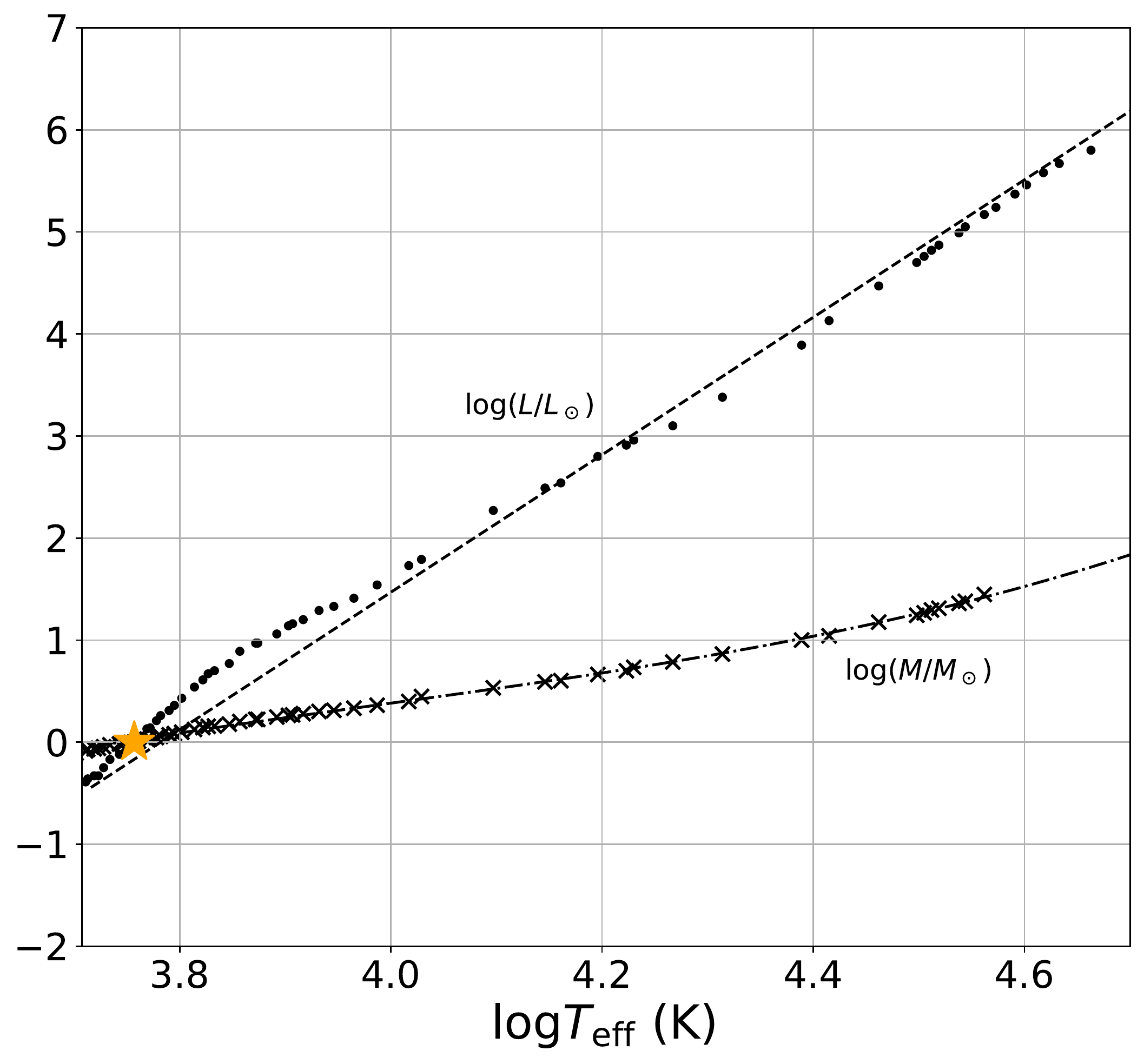}
    \caption{Main-sequence stellar parameters from \cite{Pecaut&Mamajek2013}. Black points indicate temperature-luminosity relationship, black crosses indicate temperature-mass relationship. Functions fitted to the data points are indicated by dashed and dot-dashed lines respectively. Orange star indicates the location of the Sun.}
    \label{fig: temp lumin mass relation}
\end{figure}

For the remainder of this work we use the stellar color and effective temperature sequence as in Table 5 of \cite{Pecaut&Mamajek2013}\footnote{Version 2019.3.22, see also \url{https://www.pas.rochester.edu/~emamajek/EEM_dwarf_UBVIJHK_colors_Teff.txt} } to calculate mass-loss rates using the methods described in above. Figure \ref{fig: temp lumin mass relation} shows the continuous functions fit to the data from this table for both temperature-luminosity and temperature-mass relationships. These take the the form of a power-law and a third order polynomial respectively, given by
\begin{equation} \label{eq: temp-lumin relation}
    \log(L/L_\odot) = 6.73\log(T) - 25.47
\end{equation}
and
\begin{multline}
    \log(M/M_\odot) = 1.29\log(T)^3 - 15.44\log(T)^2 \\ + 63.02\log(T) - 87.23.
\end{multline}
For the purposes of this work, we assume that the wind temperature $T$ in K remains equal to the stellar effective temperature $T_{\rm eff}$. These fits were done so that mass and luminosity could be calculated for any temperature along our temperature range. These fits were only calculated for the range of data that our temperature range encompasses in order to disregard behavior at the low temperature end of the main sequence.

\subsection{Comparison of Mass-Loss Rates}\label{subsec: mass loss comparison}

\begin{figure}[hbt!]
    \centering
    \includegraphics[width=\linewidth]{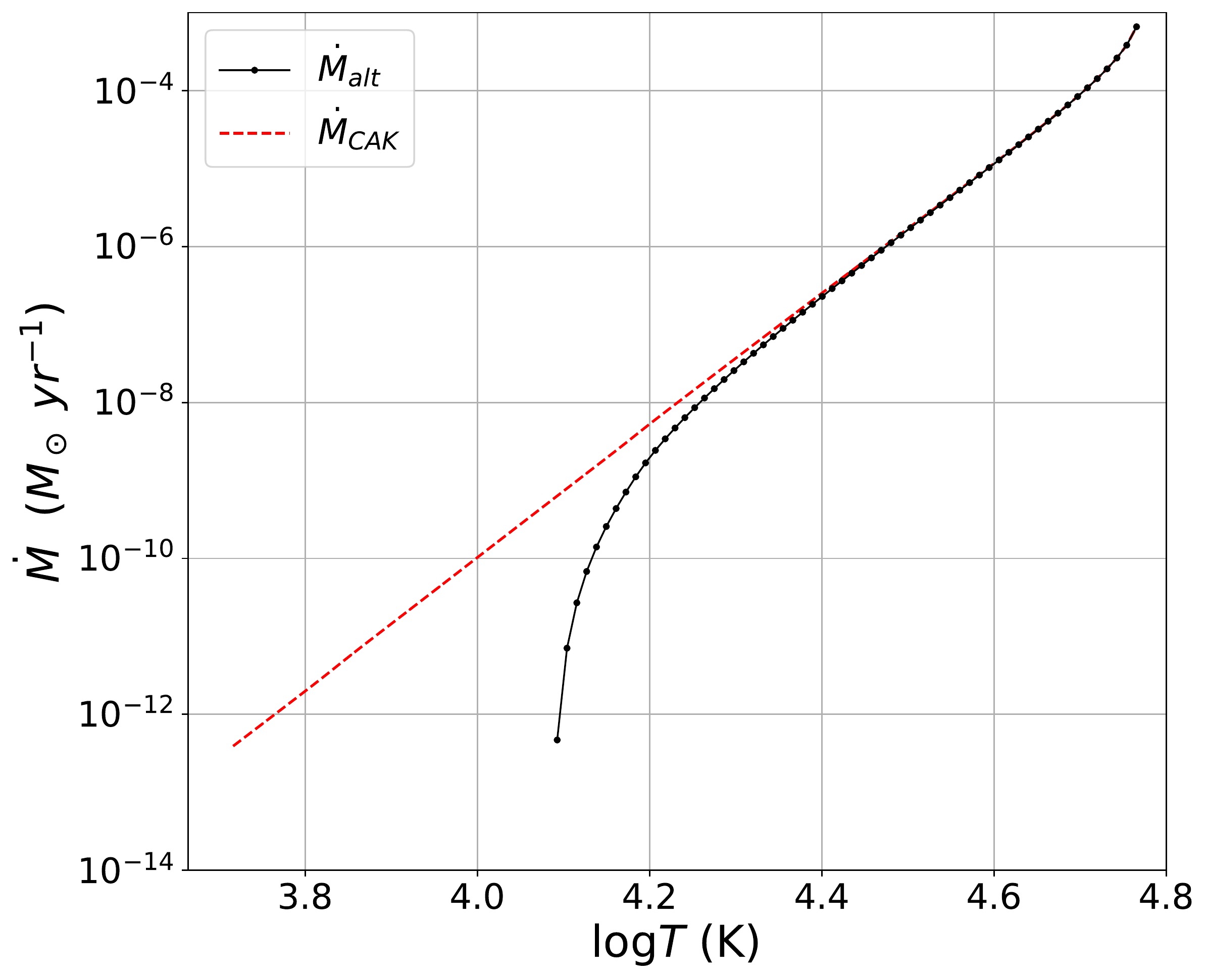}
    \caption{Preliminary comparison of $\dot{M}_{\rm CAK}$ and $\dot{M}_{\rm alt}$ for $\alpha = 0.7$, $k = 0.5$, and $\bar{Q} = 2000$. Red dashed line indicates $\dot{M}_{\rm CAK}$.}
    \label{fig: prelim mass-loss comparison}
\end{figure}
Figure \ref{fig: prelim mass-loss comparison} shows a preliminary comparison of $\dot{M}_{\rm CAK}$ and $\dot{M}_{\rm alt}$. We hold steady the parameters $\bar{Q}$, $\alpha$, $k$, and $s$, with only temperature $T$ varying, and mass and luminosity dependent on temperature as described above. There is good agreement between $\dot{M}_{\rm CAK}$ and $\dot{M}_{\rm alt}$ at high temperatures. However, there is a steep drop off exhibited at $\sim 12,000$K in $\dot{M}_{\rm alt}$, whereas $\dot{M}_{\rm CAK}$ continues as a power-law described by
\begin{equation}
    \dot{M}_{\rm CAK} \propto T^{8.51}.
\end{equation}
Using Equation (\ref{eq: temp-lumin relation}), this can also be written as 
\begin{equation} \label{eq: mass loss lumin relation}
    \dot{M}_{\rm CAK} \propto \left(\frac{L_\ast}{L_\odot}\right)^{1.26}.
\end{equation}
This is in comparison to the common form $\dot{M} \propto L_{\ast}^{1/\alpha}$. For a value of $\alpha = 0.7$ as in Figure \ref{fig: prelim mass-loss comparison}, this would take the form $\dot{M} \propto L_{\ast}^{1.42}$, whereas Equation (\ref{eq: mass loss lumin relation}) shows a slightly weaker dependence on luminosity. In Figure \ref{fig: prelim mass-loss comparison}, the black curve for $\dot{M}_{\rm alt}$ shows a strong drop-off, or quenching, which is a result of the flattening of the
force multiplier at low values of $t$.

\begin{figure}[hbt!]
    \centering
    \includegraphics[width=\linewidth]{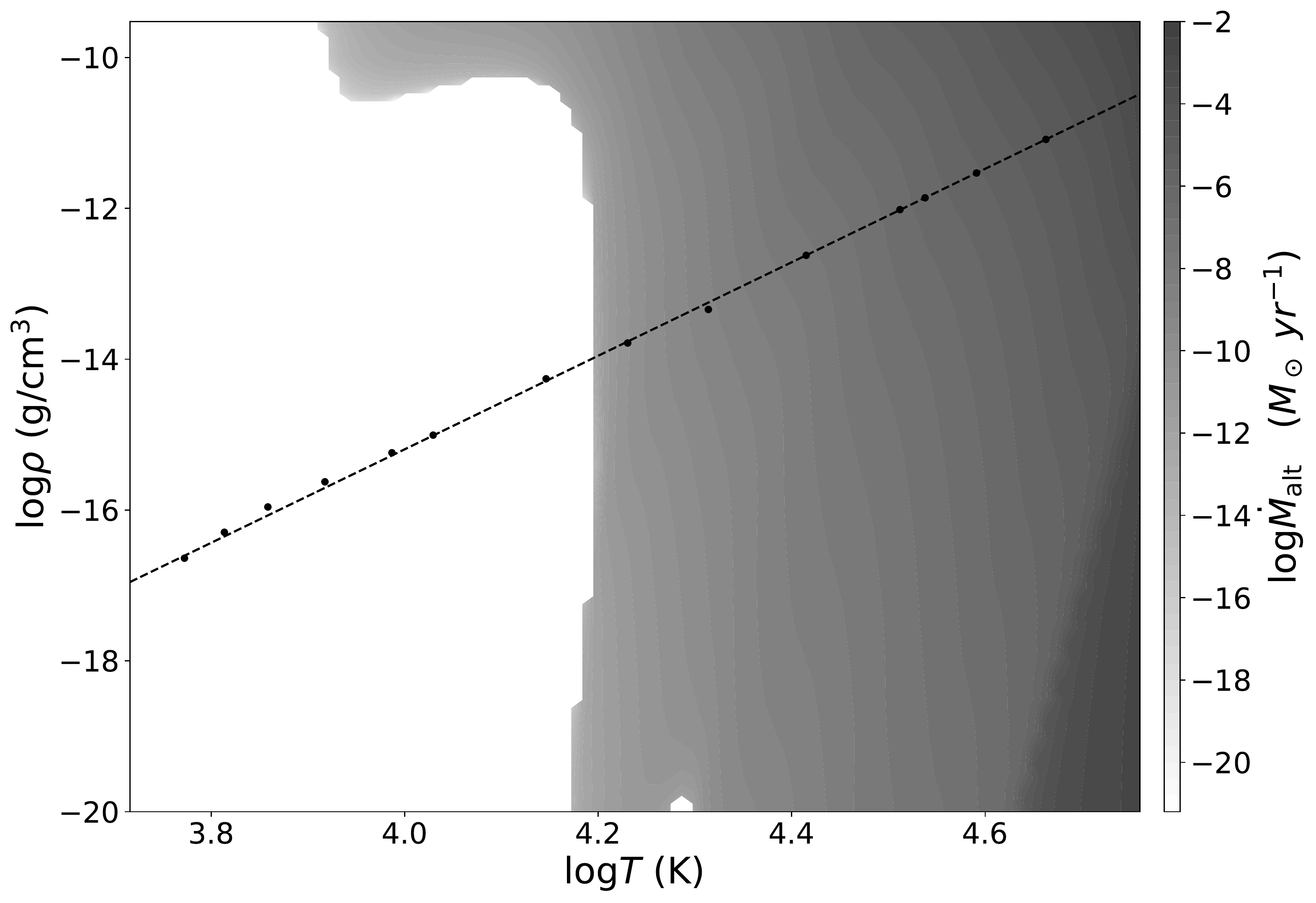}
    \caption{Contours of $\dot{M}_{\rm alt}$ over density and temperature. Dashed line indicates the CAK critical point density.}
    \label{fig: Mdot contours}
\end{figure}

\begin{figure*}[hbt!]
    \centering
    \includegraphics[width=0.9\textwidth]{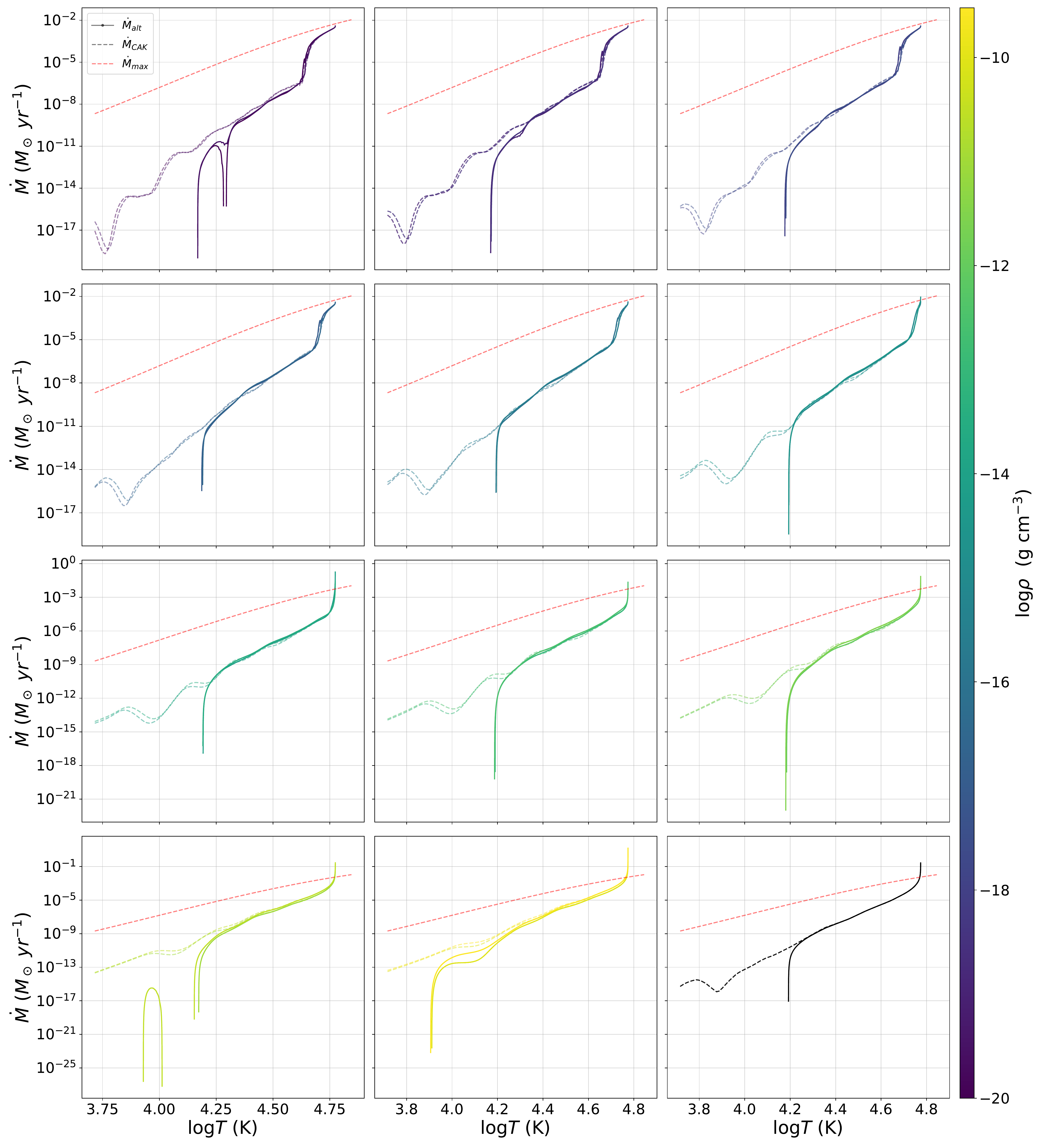}
    \caption{Comparison of $\dot{M}_{\rm CAK}$ and $\dot{M}_{\rm alt}$, by density order of magnitude. Dashed lines indicate $\dot{M}_{\rm CAK}$, solid lines indicate $\dot{M}_{\rm alt}$. Lower right corner shows calculations of $\dot{M}_{\rm CAK}$ and $\dot{M}_{\rm alt}$ for the calculated critical point densities. Red dashed lines indicate the photon-tiring limit.}
    \label{fig: CAK-Alt final comparison}
\end{figure*}

Next we introduce varying values of $\alpha$, $k$, $\bar{Q}$ and $s$. These parameters vary according to temperature and density, as seen in Section \ref{subsec: alt fit function}. 
For the remainder of this work we consider only the version of $\dot{M}_{\rm alt}$ in which all four parameters ($\alpha$, $k$, $\bar{Q}$, $s$) are allowed to vary with temperature and density. Figure \ref{fig: Mdot contours} shows the variations of $\dot{M}_{\rm alt}$ with $T$ and $\rho$.
Figure \ref{fig: CAK-Alt final comparison} compares the mass-loss rates resulting from the CAK and alternate form. As in Figure \ref{fig: prelim mass-loss comparison}, we see a sharp drop-off of $\dot{M}_{\rm alt}$ in comparison to the CAK form, commonly occurring at $\log T \approx 4.2$. At high densities ($\rho > 10^{-11}$ g~cm$^{-3}$) the departure from the CAK form is less pronounced, although still present. It should be noted that in these figures the apparent plotted end of $\dot{M}_{\rm alt}$ between $10,000$ and $20,000$~K is a result of discarding any wind solution that results in a negative wind acceleration factor $w$. Physically, this represents regions of the parameter space where the wind is quenched, a phenomenon that is not evident
when $M(t)$ is modeled as a pure power-law function of $t$. 

Also shown in Figure \ref{fig: CAK-Alt final comparison} is the photon-tiring limit, which constrains the maximum possible mass-loss rate $\dot{M}_{\rm max}$ \citep{Owocki+Gayley1997}. This is limit is defined by when the kinetic energy carried away by the wind is equal to the photon energy carried by the stellar luminosity, and is also dependent on the terminal velocity of the wind. If the terminal velocity $v_\infty$ is defined as $v_\infty = f v_{\rm esc}$, then the limit $\dot{M}_{\rm max}$ is given by
\begin{equation}
    \dot{M}_{\rm max} = \frac{2}{f^2} \frac{L_\ast}{v_{\rm esc}^2},
\end{equation}
with the escape velocity given by $v_{\rm esc}^2 = 2GM_\ast/R_\ast$. For this work we take the traditional value of $f=3$. For lower densities ($\rho \lesssim 10^{-16}$ g/cm$^{-3}$), we see that the photon-tiring limit intersects the mass-loss rate curves at approximately the same point at high temperatures where $\Gamma \geq 1$.

\section{Discussion and Conclusions}\label{sec:Conclusions}
In this work, we have constructed an updated list of atomic data, with data for 4,514,900 spectral lines taken from the NIST, CMFGEN, CHIANTI, and TOPbase databases. These atomic data were then used to calculate the line strength parameter $q_i$ for each line for a density range of $10^{-20}$ to $10^{-10}$ g~cm$^{-3}$ over a temperature range of 5,200 to 70,000 K\footnote{Atomic data and other parameters calculated in the course of this work are available at \url{https://github.com/aslyv2/Rad-Winds}}. The weighting function $\widetilde{W_i}$ was also calculated. These parameters were used to find the line force multiplier $M(t)$ over a range of $t$ from $10^{-15}$ to $10$. The distribution of $M(t)$ was fit using a power-law as described by Equation (\ref{eq:complete CAK M(t)}), as well as an alternate function in the form of a saturated power-law, as described by  Equation (\ref{eq: Alt M(t)}).
We found that this alternative function better describes the values of the line-force multiplier as calculated from the updated line list, especially at low values of $t$. The residuals of this alternate function are consistently lower than those that result from the CAK form in the case of low $t$, and comparable for high values of $t$. This is consistent across temperatures and densities. We include the parameter $s$ to control the sharpness of the turn-over from the power-law segment to the flat segment. $M_{\rm alt}(t)$ reduces to the power-law form in the limit of high-$t$ for all values of the sharpness parameter s. In the limit of low-$t$ $M_{\rm alt}(t)$ similarly reduces to the calculated value of $\bar{Q}$.

Using the alternate function for $M(t)$, we also calculate mass-loss rates for the temperatures and densities in our grid, using the fitted parameters $\alpha, k, \textrm{ and } s$, along with the corresponding calculated values of $\bar{Q}$. We find that the sharpness parameter $s$ has a drastic effect on the determined mass-loss rates, especially at high temperatures. Additionally, there is a sharp drop-off in the mass-loss rates calculated from the updated form of $M(t)$ and a resulting discrepancy between it and the CAK mass-loss form. This drop-off in the mass-loss rate describes a quenching of the line-driven wind that is not present in the CAK form.

We find that the quenching of the wind typically occurs between temperatures of $10,000$ K and $20,000$ K and at luminosities of ${2.5} \lesssim \log(L_*/L_\odot) \lesssim {2.75}$. This may be a partial explanation for the discrepancy noted between empirically derived mass-loss rates and predicted values for stars of luminosities below $\sim 10^5 \, L_{\odot}$ ($T\approx 36,000$ K), often referred to in the literature as the ``weak-wind problem'' \citep{Muijres2012}, although it should be noted that our calculations place the quenching of the wind at lower luminosities and temperatures. 
It is also possible that these effects could be important to include when modeling the oscillations of Slowly Pulsating B (SPB) stars, which have $T_{\rm eff}$ values between about 10,000 and 20,000 K \citep{DeCat2007}. The interactions between their oscillations and winds remain poorly understood \citep[e.g.,][]{Saio2015}.

Lastly, there is another physical effect that must be taken into account to fully understand how the predicted quenching effect manifests itself: collisionless decoupling between the line-driven ions and the dominant hydrogen/helium gas. This has been proposed to be important both for B-type stars \citep{Springmann&Pauldrach1992,Babel1996,Krticka2014} and some metal-enriched AGNs \citep{Baskin2012}.
In some low-density systems this decoupling can lead to frictional heating with wind temperatures far in excess of the stellar $T_{\rm eff}$, and in others may produce fully-separated multi-component winds with peculiar chemical abundances. It may be possible that the drastic reduction in the ion line-force (which arises due to the flattening of the force multiplier) allows these systems to undergo a more gentle and gradual transition from a coupled single-fluid outflow to a quiescent hydrostatic atmosphere.

Although here we consider only the assumption of LTE, other similar works consider the effects of NLTE (see, for example, \citealp{Gormaz-Matamala2019}). \cite{Puls2000} accounted for NLTE effects in the line distribution by restricting the types of lines used to those with or directly connected to those with a ground or metastable lower level. For our purposes, it will be useful to refine the ionization balance used here using the modified nebular approach described by others (see, for example, \citealp{Abbott+Lucy1985}, \citealp{Gormaz-Matamala2019}). Although the assumption of a Planck function for $F(\nu)$ allowed us to maintain generality in this work, in future work it will be necessary to refine our choice of $F(\nu)$ to a more realistic distribution. For example, a self-consistent treatment of absorption in the near-star atmosphere could be applied to the phenomenon of ``bistability'' \citep[e.g][]{Lamers1991} in which the wind sees a lower flux shortward of 91.2 nm---and a higher flux in the Balmer continuum---and this affects the relative strengths of the lines that contribute to $M(t)$. Alternatively, this bistability jump could be a result of the recombination of Fe between Fe~III and Fe~IV, with the contribution of the Fe~III lines dominating the radiative acceleration of the subsonic part of the wind \citep{Vink1999,Vink2000}. \cite{Puls2000} similarly found that at low line strengths mass-loss is dominated largely by the radiative acceleration of iron group elements, with lighter ions playing a more important role at larger line strength. This bistability jump is predicted to be reflected by an increase in mass loss, occurring around $\sim$20,000 K. Our calculations show that the relevant recombination temperature of Fe~IV does occur at a local maximum of the CAK $k$ parameter (see Figure \ref{fig: Alt parameter evolution}(a)). However, while the mass-loss rate is usually quite sensitive to $k$, we do not see any significant increase in our final calculations for $\dot{M}$ at these parameters (e.g., Figure \ref{fig: CAK-Alt final comparison}).

In this study we have also limited ourselves to the solar elemental abundances of \citet{Asplund2009}. Other abundance patterns, such as those found in nearby galaxies with lower metallicity \citep{Puls+1996} or in certain types of chemically peculiar stars \citep{Alecian2019} should be explored. Additionally, we plan to explore the radial dependence of the $t$ parameter and the associated spatial variation of $M(t)$ in self-consistent models of radial outflow from stars and other luminous astrophysical sources such as active galactic nuclei.

\acknowledgments
CHIANTI is a collaborative project involving George Mason University, the University of Michigan (USA), University of Cambridge (UK) and NASA Goddard Space Flight Center (USA). This work was supported by start-up funds from the Department of Astrophysical and Planetary Sciences at the University of Colorado Boulder. This research made use of NASA's
 Astrophysics Data System (ADS). The authors would also like to thank the anonymous referee for their helpful comments.

\software{Python v3.7.6 \citep{python},
NumPy \citep{numpy1,numpy2},
SciPy \citep{scipy},
matplotlib \citep{matplotlib},
AstroPy \citep{astropy1,astropy2}}

\pagebreak
\appendix 
\restartappendixnumbering
\section{Database Selection and Specific Line Counts}\label{database appendix}

In cases where more than one database listed atomic data for a given ion, the database with the largest number of available transitions was used for each ionization state of each element. In general, CMFGEN was found to contain the most lines for a majority of ions. However, where CMFGEN data was nonexistent or insufficient, the database with the next most lines was used. Generally, this was CHIANTI. For several elements and ionization states (namely N VI, N VII, Cl VIII, Cl IX, and Ni X), the necessary atomic data was not available from the databases used. Table \ref{Table: database by ion} shows a breakdown of line counts $n$ by ion, with the database used for each also listed. In some databases, transitions with a both a radiative decay rate and an autoionization rate were represented twice. After compiling the line list from the total available data, we discarded any such duplicate transitions. 

\startlongtable
\begin{deluxetable*}{llllllllllll}
\tabletypesize{\footnotesize}
\tablecolumns{12}

\tablewidth{\textwidth}
\tablecaption{Number of lines ($n$) and database used for each ion. A dash (-) indicates that no data were available. CMFGEN, NIST, CHIANTI, and TOPbase are abbreviated as CM, N, CH, and T respectively.}
\label{Table: database by ion}

\tablehead{
\colhead{Ion} & \colhead{$n$} & \colhead{Database} & \colhead{Ion}  & \colhead{$n$} & \colhead{Database} & \colhead{Ion} & \colhead{$n$}  & \colhead{Database} & \colhead{Ion} & \colhead{$n$} & \colhead{Database}}

\startdata
H I       & 435            &   CM   &    Na V      &    10644          &   CM   &    Cl IV     & 8612           &   CM   &    V III     & 21             &  N      \\   
He I      & 3857           &   CM   &    Na VI     &    10994          &   CM   &    Cl V      & 3388           &   CM   &    V IV      & 239            &  N      \\   
He II     & 435            &   CM   &    Na VII    &    5436           &   CM   &    Cl VI     & 2377           &   CM   &    V V       & 10             &  N      \\   
Li I      & 68             &   N    &    Na VIII   &    4742           &   CM   &    Cl VII    & 1557           &   CM   &    V VI      & 4              &  N      \\   
Li II     & 134            &   N    &    Na IX     &    4201           &   CH   &    Cl VIII   & -              &   -    &    V VII     & 7              &  N      \\      
Li III    & 2              &   N    &    Na X      &    331            &   CH   &    Cl IX     & -              &   -    &    V VIII    & 6              &  N      \\     
Be I      & 175            &   N    &    Mg I      &    2841           &   CM   &    Cl X      & 24             &   CH   &    V IX      & 16             &  N      \\      
Be II     & 97             &   N    &    Mg II     &    2641           &   CH   &    Ar I      & 3824           &   CM   &    V X       & 13             &  N      \\    
Be III    & 100            &   N    &    Mg III    &    4753           &   CH   &    Ar II     & 79388          &   CM   &    Cr I      & 49885          &  CM     \\     
Be IV     & 10             &   N    &    Mg IV     &    5706           &   CM   &    Ar III    & 6901           &   CM   &    Cr II     & 66400          &  CM     \\     
B I       & 96             &   N    &    Mg V      &    6377           &   CM   &    Ar IV     & 11290          &   CM   &    Cr III    & -              &  -      \\    
B II      & 150            &   N    &    Mg VI     &    14480          &   CM   &    Ar V      & 8350           &   CM   &    Cr IV     & 67061          &  CM     \\     
B III     & 74             &   N    &    Mg VII    &    11940          &   CM   &    Ar VI     & 5              &   N    &    Cr V      & 43860          &  CM     \\     
B IV      & 134            &   N    &    Mg VIII   &    5820           &   CM   &    Ar VII    & 35             &   CH   &    Cr VI     & 4406           &  CM     \\     
B V       & 58             &   N    &    Mg IX     &    5517           &   CM   &    Ar VIII   & 2743           &   CH   &    Cr VII    & 46             &  CH     \\   
C I       & 10204          &   CM   &    Mg X      &    26078          &   CH   &    Ar IX     & 5691           &   CH   &    Cr VIII   & 131            &  CH     \\   
C II      & 8017           &   CM   &    Al I      &    4985           &   CM   &    Ar X      & 4435           &   CH   &    Cr IX     & 236            &  CH     \\     
C III     & 9468           &   CM   &    Al II     &    2870           &   CM   &    K I       & 1471           &   CM   &    Cr X      & 16             &  N      \\     
C IV      & 1297           &   CM   &    Al III    &    2665           &   CM   &    K II      & 38603          &   CM   &    Mn I      & 164            &  N      \\   
C V       & 2196           &   CM   &    Al IV     &    5296           &   CH   &    K III     & 220            &   CM   &    Mn II     & 49066          &  CM     \\     
C VI      & 1575           &   CM   &    Al V      &    6607           &   CH   &    K IV      & 18227          &   CM   &    Mn III    & 70218          &  CM     \\   
N I       & 855            &   CM   &    Al VI     &    7989           &   CM   &    K V       & 7252           &   CM   &    Mn IV     & 72374          &  CM     \\     
N II      & 7879           &   CM   &    Al VII    &    15486          &   CM   &    K VI      & 14870          &   CM   &    Mn V      & 77009          &  CM     \\   
N III     & 6710           &   CM   &    Al VIII   &    13501          &   CM   &    K VII     & 71             &   N    &    Mn VI     & 70116          &  CM     \\    
N IV      & 13886          &   CM   &    Al IX     &    5859           &   CM   &    K VIII    & 95             &   N    &    Mn VII    & 8277           &  CM     \\    
N V       & 1296           &   CM   &    Al X      &    5041           &   CM   &    K IX      & 2758           &   CH   &    Mn VIII   & 47             &  CH     \\  
N VI      & 2263           &   CM   &    Si I      &    2791           &   CM   &    K X       & 5700           &   CH   &    Mn IX     & 137            &  CH     \\   
N VII     & 3150           &   CM   &    Si II     &    4196           &   CM   &    Ca I      & 106            &   N    &    Mn X      & 236            &  CH     \\   
O I       & 4145           &   CM   &    Si III    &    1328           &   CM   &    Ca II     & 238            &   CH   &    Fe I      & 141928         &  CM     \\   
O II      & 17874          &   CM   &    Si IV     &    2672           &   CH   &    Ca III    & 520            &   N    &    Fe II     & 530827         &  CM     \\     
O III     & 6516           &   CM   &    Si V      &    5354           &   CH   &    Ca IV     & 2              &   N    &    Fe III    & 136060         &  CM     \\   
O IV      & 7599           &   CM   &    Si VI     &    6518           &   CM   &    Ca V      & 9              &   CH  &    Fe IV     & 72223          &  CM      \\  
O V       & 3237           &   CM   &    Si VII    &    9364           &   CM   &    Ca VI     & 10             &   CH  &    Fe V      & 71983          &  CM      \\  
O VI      & 1569           &   CM   &    Si VIII   &    705            &   CH   &    Ca VII    & 86             &   CH  &    Fe VI     & 185392         &  CM      \\    
O VII     & 3505           &   CM   &    Si IX     &    403            &   CH   &    Ca VIII   & 200            &   CH  &    Fe VII    & 86504          &  CM      \\   
O VIII    & 1575           &   CM   &    Si X      &    5017           &   CH   &    Ca IX     & 9230           &   CH  &    Fe VIII   & 21134          &  CH      \\     
F I       & 119            &   N    &    P I       &    46             &   N    &    Ca X     & 2760            &  CH   &    Fe IX      & 47085          &  CH     \\   
F II      & 2354           &   CM   &    P II      &    217043         &   CM   &    Sc I      & 259            &  N    &    Fe X      & 50854          &  CM      \\    
F III     & 9725           &   CM   &    P III     &    5576           &   CM   &    Sc II     & 77253          &  CM   &    Co I      & 118            &  N       \\      
F IV      & 15             &   N    &    P IV      &    2537           &   CM   &    Sc III    & 687            &  CM   &    Co II     & 61986          &  CM      \\      
F V       & 11             &   N    &    P V       &    2700           &   CH  &    Sc IV     & 4              &  N     &    Co III    & 679412         &  CM      \\     
F VI      & 8415           &   T    &    P VI      &    5533           &   CH  &    Sc V      & 4              &  N     &    Co IV     & 69425          &  CM      \\    
F VII     & 6406           &   T    &    P VII     &    3              &   CH  &    Sc VI     & -              &  -     &    Co V      & 75923          &  CM      \\     
F VIII    & 5614           &   T    &    P VIII    &    25             &   CH  &    Sc VII    & 15             &  N     &    Co VI     & 75118          &  CM      \\   
F IX      & 3488           &   T    &    P IX      &    59             &   CH  &    Sc VIII   & 16             &  N     &    Co VII    & 68388          &  CM      \\   
Ne I      & 2629           &   CM   &    P X       &    78             &   CH  &    Sc IX     & 15             &  N     &    Co VIII   & 88548          &  CM      \\    
Ne II     & 5795           &   CM   &    S I       &    19813          &   CM  &    Sc X      & -              &  -     &    Co IX     & 12232          &  CM      \\   
Ne III    & 2343           &   CM   &    S II      &    8527           &   CM  &    Ti I      & 490            &  N     &    Co X      & 5              &  N       \\      
Ne IV     & 9725           &   CM   &    S III     &    4543           &   CM  &    Ti II     & 93118          &  CM    &    Ni I      & 188            &  N       \\   
Ne V      & 13037          &   CM   &    S IV      &    7530           &   CM  &    Ti III    & 21722          &  CM    &    Ni II     & 51812          &  CM      \\  
Ne VI     & 5171           &   CM   &    S V       &    3605           &   CM  &    Ti IV     & 1000           &  CM    &    Ni III    & 66511          &  CM      \\     
Ne VII    & 5213           &   CM   &    S VI      &    1936           &   CM  &    Ti V      & 4              &  N     &    Ni IV     & 72898          &  CM      \\      
Ne VIII   & 26832          &   CH   &    S VII     &    73             &   N   &    Ti VI     & 11             &  N     &    Ni V      & 75541          &  CM      \\     
Ne IX     & 216            &   CH   &    S VIII    &    54             &   N   &    Ti VII    & 1              &  N     &    Ni VI     & 79169          &  CM      \\      
Ne X      & 190            &   CH   &    S IX      &    51             &   N   &    Ti VIII   & 15             &  N     &    Ni VII    & 74411          &  CM      \\      
Na I      & 2778           &   CM   &    S X       &    57             &   N   &    Ti IX     & 14             &  N     &    Ni VIII   & 71614          &  CM      \\      
Na II     & 5054           &   CH  &    Cl I       &    75             &   N   &    Ti X      & 43             &  N     &    Ni IX     & 79227          &  CM      \\      
Na III    & 4368           &   CH  &    Cl II      &    52             &   N   &    V I       & 1095           &  N     &    Ni X      & -              &  CM      \\     
Na IV     & 3754           &   CM  &    Cl III     &    50             &   N   &    V II      & 1415           &  N     &    -         & -              &  -       \\
\enddata

\end{deluxetable*}

\bibliography{rad_winds_bibliography.bib}
\bibliographystyle{aasjournal}

\end{document}